\documentclass[10pt,prl,aps,twocolumn,superscriptaddress,floatfix]{revtex4-1}

\pdfoutput=1

\usepackage{graphicx}
\usepackage{amssymb}
\usepackage{bm}
\usepackage{color}

\begin{document}

\title{Critical level crossings and gapless spin liquid \\ in the square-lattice spin-$1/2$ $J_1$-$J_2$ Heisenberg antiferromagnet}

\author{Ling Wang}
\email{lingwang@csrc.ac.cn}
\affiliation{Beijing Computational Science Research Center, 10 East Xibeiwang Rd, Beijing 100193, China}

\author{Anders W. Sandvik}
\email{sandvik@bu.edu}
\affiliation{Department of Physics, Boston University, 590 Commonwealth Ave, Boston, Massachusetts 02215, USA}
\affiliation{Beijing Computational Science Research Center, 10 East Xibeiwang Rd, Beijing 100193, China}
\affiliation{Beijing National Laboratory of Condensed Matter Physics and Institute of Physics, Chinese Academy of Sciences, Beijing 100190, China}

\date{\today}

\begin{abstract}
  We use the DMRG method to calculate several energy eigenvalues of the
  frustrated $S=1/2$ square-lattice $J_1$-$J_2$ Heisenberg model on
  $2L \times L$ cylinders with $L \le 10$. We identify excited-level
  crossings versus the coupling ratio $g=J_2/J_1$ and study their drifts with
  the system size $L$. The lowest singlet-triplet and singlet-quintuplet
  crossings converge rapidly (with corrections $\propto L^{-2}$) to different
  $g$ values, and we argue that these correspond to ground-state transitions
  between the N\'eel antiferromagnet and a gapless spin liquid, at
  $g_{c1} \approx 0.46$, and between the spin liquid and a valence-bond-solid
  at $g_{c2} \approx 0.52$. Previous studies of order parameters were not
  able to positively discriminate between an extended spin liquid phase and a
  critical point. We expect level-crossing analysis to be a generically
  powerful tool in DMRG studies of quantum phase transitions.
\end{abstract}
 
\maketitle

The spin-$1/2$ frustrated $J_1$-$J_2$ Heisenberg model on the two-dimensional (2D)
square lattice (where $J_1$ and $J_2$ are the strengths of the first and second 
neighbor couplings ${\bf S}_i\cdot {\bf S}_j$, respectively) has been studied and debated since the early days 
of the high-$T_c$ cuprate superconductors
\cite{chandra88,ed1,gelfand89,sachdev89,ed2,singh90,RS9173,ed3,ivanov92,ed5,ed4,Singh_prb60.7278}.
The initial interest in the system stemmed from the proposal that frustrated
antiferromagnetic (AFM) couplings could lead to a spin liquid (SL) in which
preformed pairs (resonating valence bonds \cite{fazekas74}) become
superconducting upon doping \cite{RVB,PatrickRVB}. Later, with frustrated
quantum magnets emerging in their own right as an active research field
\cite{diep05}, the $J_1$-$J_2$ model became a prototypical 2D system for
theoretical and computational studies of quantum phase transitions and
nonmagnetic
states~\cite{Capriotti_prl84.3173,Capriotti_prl87.097201,mambrini_prb74.144422,sirker_prb73.184420,darradi_prb78.214415,WangJ1J216,arlego_prb78.224415,Beach_prb79.224431,richter_ed,HuJ1J2,JiangJ1J2,ShengJ1J2,MurgJ1J2,KaoJ1J2,WangJ1J2,mambrini17,HaghshenasJ1J2}. Of
primary interest is the transition from the long-range N\'eel AFM ground
state \cite{AndersonTower,chakravarty89,manousakis91} at small $g=J_2/J_1$ 
to a nonmagnetic state in a window around $g\approx 0.5$ (before a stripe AFM phase at
$g \agt 0.6$). The nature of this quantum phase transition has remained
enigmatic~\cite{Singh_prb60.7278,Capriotti_prl84.3173,sirker_prb73.184420,darradi_prb78.214415,WangJ1J216,mambrini17},
despite a large number of calculations with numerical tools of ever
increasing sophistication, e.g., the density matrix renormalization group
(DMRG) method \cite{whitedmrg,schollwoechreview,JiangJ1J2,ShengJ1J2},
tensor-product states~\cite{MurgJ1J2,KaoJ1J2,WangJ1J2,WangJ1J216,mambrini17,HaghshenasJ1J2},
and variational Monte Carlo \cite{HuJ1J2,Imada15}.

The nonmagnetic state may be one with spontaneously broken lattice symmetries due 
to formation of a pattern of singlets (a valence-bond-solid, VBS) or a SL. Within these 
two classes of potential ground states there are several different proposals, e.g., 
a columnar \cite{RS9173,singh90,Singh_prb60.7278} versus a
plaquette~\cite{Capriotti_prl84.3173,mambrini_prb74.144422,ShengJ1J2,KaoJ1J2}
VBS, and gapless~\cite{HuJ1J2} or gapped~\cite{JiangJ1J2} SLs. The quantum
phase transition out of the AFM state may possibly be an unconventional
'deconfined' transition \cite{senthil04,DQCP2,moon12}, which recently has
been investigated primarily within other models
\cite{DQCP3,melko08,lou09,banerjee10,block13,harada13,chen13,shao16,nahum15}
hosting direct AFM--VBS transitions. In the $J_1$-$J_2$ model, some studies
have indicated that the nonmagnetic phase may actually comprise two different
phases, with an entire gapless SL phase---not just a critical
point---existing between the AFM and VBS states
\cite{ShengJ1J2,Imada15}. However, because of the small system sizes
accessible, it was not possible to rule out a direct AFM--VBS transitions. We
here demonstrate an intervening gapless SL by locating the AFM--SL and SL-VBS
transitions using a numerical level-spectroscopy approach, where finite-size
transition points are defined using excited-level crossings. These crossing
points exhibit smooth size dependence and can be more reliably extrapolated
to infinite size than the order parameters and gaps used in past studies.

We use a variant of the DMRG
method~\cite{whitedmrg,whiteparadmrg,McCulloch07,schollwoechreview} to
calculate the ground state energy as well as several of the lowest singlet,
triplet and quintuplet excited energies. In the AFM state, the lowest
excitation above the singlet ground state in a finite system with an even
number of sites is a triplet---the lowest state in the Anderson tower of
'quantum rotor' states~\cite{AndersonTower}. If the nonmagnetic ground state
is a degenerate singlet when the system length $L\to \infty$, as it should be
in both a VBS and a topological (gapped) SL, there must be a crossing of the
lowest singlet and triplet excitation at a point $g(L)$ that approaches $g_c$
with increasing $L$. This is indeed observed at the dimerization transition
of the 1D $J_1$-$J_2$ chain \cite{nomura92,eggert96,sandvik10a} and related
systems \cite{sandvik10b,suwa_prl15}, and size extrapolations
give $g_c$ to remarkable precision, even with system sizes only up to
$L \approx 30$. A level crossing with the same finite-size behavior was
observed recently also in the 2D $J$-$Q$ model~\cite{suwa_prb94.144416},
which is a Heisenberg model supplemented by four-spin interactions causing
an AFM--VBS transition~\cite{DQCP3,melko08,lou09,banerjee10,block13,harada13,chen13},
likely a deconfined quantum-critical point with unusual scaling properties
\cite{shao16}.  It is then natural to investigate level crossings also in the
2D $J_1$-$J_2$ model.

We will demonstrate a singlet-triplet level crossing in the $J_1$-$J_2$ model
which for $2L \times L$ cylindrical lattices shifts as
$g_{c2}-g_{c2}(L) \propto L^{-2}$ and converges to $g_{c2} \approx 0.52$. We
also observe a singlet-quintuplet level crossing, which
converges to a different point, $g_{c1} \approx 0.46$. Given the known
transitions associated with singlet-triplet crossings, and that a
singlet-quintuplet crossing was found at the transition between the critical
and AFM states in a Heisenberg chain with long-range interactions
\cite{sandvik10a,sandvik10anote}, we interpret both $g_{c1}$ and $g_{c2}$ as
quantum-critical points. For $g_{c1} \le g \le g_{c2}$ the system appears
to be a gapless SL with algebraically decaying correlations, as in one of the 
scenarios proposed in Refs.~\onlinecite{ShengJ1J2,Imada15} (and previously 
discussed also in Ref.~\onlinecite{sandvik12}). Our value of $g_{c1}$ is
in the middle of the range $g=0.4 \sim 0.5$ where most recent studies have
put the end of the AFM phase \cite{JiangJ1J2,HuJ1J2,ShengJ1J2,Imada15}, and
$g_{c2}$ is close to the VBS-ordering point in
Refs.~\onlinecite{ShengJ1J2,Imada15}.

\begin{figure}
\begin{center}
\includegraphics[width=7cm]{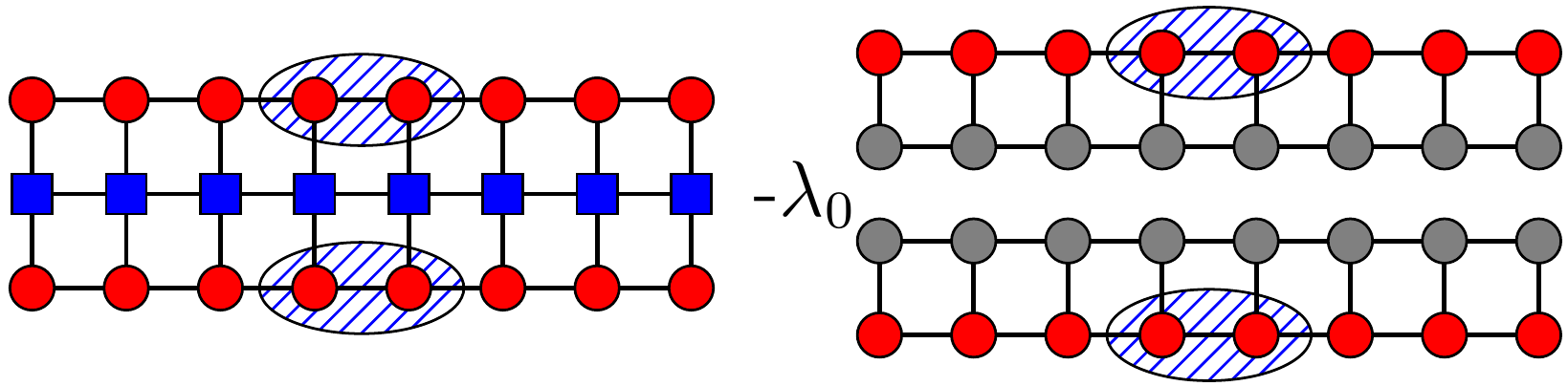}
\vskip-1mm
\caption{Illustration of the effective Hamiltonian $H_{\text{eff}}^1$ below in
  Eq.~(\ref{Heff}).  Red and gray circles represent the targeted state
  $|\psi_1\rangle$ and the ground state $|\psi_0\rangle$, respectively, and
  the blue squares show the original Hamiltonian as a matrix-product
  operator. The hatched area represents $U_1^{\dagger}|\psi_1\rangle$, where
  $U_1$ projects to the canonical MPS for $|\psi_1\rangle$ without the
  hatched area.}
\label{dmrg_demo}
\end{center}
\vskip-2mm
\end{figure}

{\it DMRG calculations.---}The DMRG method \cite{whitedmrg} is a powerful
tool for computing the ground state $|\psi_0\rangle$ of a many-body
Hamiltonian. By solving a Hamiltonian $H_{\text{eff}}$ in a relevant
low-entangled subspace of the full Hilbert space, one can obtain an effective
wavefunction, through which the most relevant subspace is selected for the
next iteration. A series of such subspace projectors produces the ground
state as a matrix product state (MPS), i.e., the wavefunction coefficients
are traces of products of local matrices of chosen size
$m$~\cite{ostlund95,schollwoechreview}.

The lowest excited state $|\psi_1\rangle$ can also be targeted with DMRG
\cite{McCulloch07} provided that $|\psi_0\rangle$ has been pre-calculated.
The only difference from a ground-state DMRG algorithm is that one has
to maintain the orthogonality condition $\langle\psi_1|\psi_0\rangle=0$ at
each step. Upon reformulating the Hamiltonian for the lowest excited
state as $H_1=H-\lambda_0|\psi_0\rangle\langle\psi_0|$, where $\lambda_0$ is
the eigenvalue of $H$ corresponding to $|\psi_0\rangle$, one can write down the 
effective Hamiltonian equation in the DMRG procedure as
\begin{equation}
\label{Heff}
\left[U_1^{\dagger}\left(H-\lambda_0|\psi_0\rangle\langle\psi_0|\right)U_1\right]
U_1^{\dagger}|\psi_1\rangle=\lambda_1U_1^{\dagger}|\psi_1\rangle,
\end{equation}
where $U_1$ projects onto the canonical MPS~\cite{schollwoechreview} for $|\psi_1\rangle$
without the center two sites, as illustrated in Fig.~\ref{dmrg_demo}, and $\lambda_1$ is
the eigenvalue for $|\psi_1\rangle$. We can therefore define an effective Hamiltonian
$H^1_{\text{eff}}\equiv U_1^{\dagger}\left(H-\lambda_0|\psi_0\rangle\langle\psi_0|\right)U_1$.

Similarly, given that $|\psi_i\rangle$ for all $i<j$ ($\lambda_i<\lambda_j$)
have been pre-calculated, we observe that one can compute the next eigenstate
$j$ as an MPS with a given number of kept Schmidt states $m$ using a
modified Hamiltonian
\begin{equation}
H_j=H-\sum_{i=0}^{j-1}\lambda_i|\psi_i\rangle\langle\psi_i|.
\end{equation}
Here
$H_{\text{eff}}^jU_j^{\dagger}|\psi_j\rangle=\lambda_jU_j^{\dagger}|\psi_j\rangle$
as in Eq.~(\ref{Heff}). In practice such a DMRG scheme will break down (i.e.,
unreasonably large $m$ has to be used) when the eigenstates far from the
bottom of the spectrum begin to violate the area law.

The $2L\times L$ cylinder geometry, with open and periodic boundaries in the $x$ and $y$
direction, respectively, is known to be suitable for 2D DMRG calculations \cite{white07} 
and we use it here for even $L$ up to $10$. We employ the DMRG with either U(1) (the total 
spin $z$ component $S^z$ is conserved) or SU(2) symmetry. With $U(1)$ symmetry, we
generate up to ten $S^z=0$ states and obtain the total spin $S$ by 
computing the expectation value of ${\bf S}^2$.

An advantage of focusing on the level spectrum is the well known fact that
the energy converges much faster with the number $m$ of Schmidt states than
other physical observables, and also as a function of the number of sweeps in
the DMRG procedure. We here apply very stringent convergence criteria and
also extrapolate away the remaining finite-$m$ errors based on calculations
for several values of $m$ up to $m=12000$ with $U(1)$ symmetry and $m=5000$
with $SU(2)$ symmetry.  The DMRG procedures and extrapolations are further
discussed in Supplemental Material (SM) \cite{sm}.

\begin{figure}[t]
\begin{center}
\includegraphics[width=8.0cm]{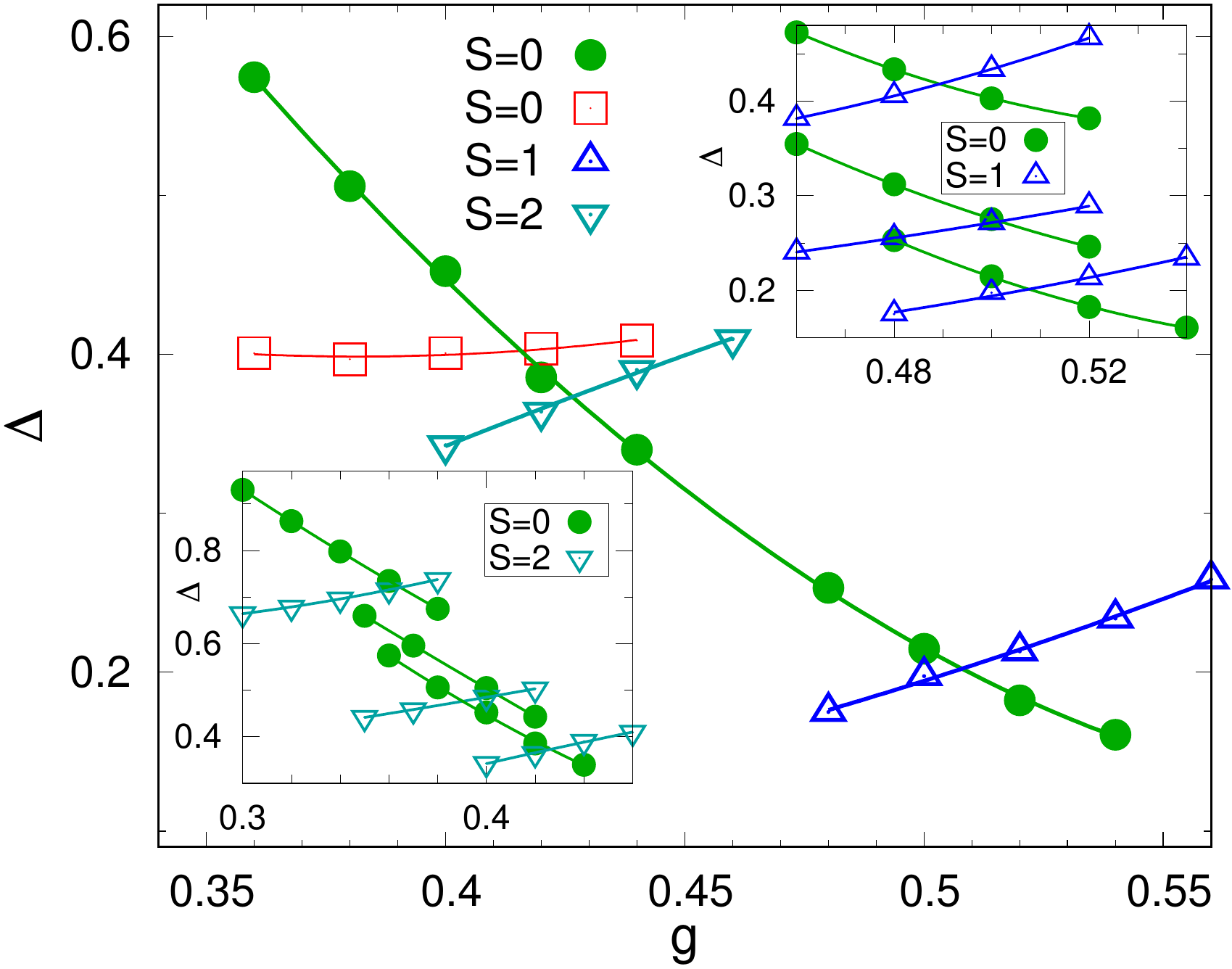}
\vskip-1mm
\caption{Gaps to the relevant $S=0,1$, and $2$ excitations vs $g$ for $L=10$. The insets show 
the regions of the level crossings of interest for $L=6,8,10$ (gaps decreasing with increasing 
$L$). The curves show polynomial fits.}
\label{gapplot}
\end{center}
\vskip-3mm
\end{figure}

{\it Results.---}Figure \ref{gapplot} shows two singlet gaps and the lowest
triplet and quintuplet gaps versus $g$ in and close to the non-magnetic
regime. The main graph shows results for $L=10$. One of the singlet gaps
decreases rapidly with increasing $g$, crossing the other three levels. This
is the lowest singlet excitation starting from $g\approx 0.42$, after
crossing the other singlet (which has other quantum numbers related to the
lattice symmetries) that is lower in what we will argue is the AFM phase. The
insets of Fig.~\ref{gapplot} show results also for $L=6$ and $8$ in the
region around the level crossings that we will analyze (the higher gaps for
$L=4$ are not shown for clarity). Using polynomial fits to the DMRG data
points, we extract crossing points $g_{c1}(L)$ between the singlet and the
quintuplet, as well as $g_{c2}(L)$ between the singlet and the triplet. The
singlet-singlet crossings taking place close to $g_{c1}(L)$ are discussed in
the SM \cite{sm}; their size dependence is similar to $g_{c1}(L)$. For
$g \agt g_{c1}(L)$ there are also other levels in the energy range of
Fig.~\ref{gapplot}, including singlets, but the $S=0,1,2$ gaps graphed are
the lowest with these spins up to and beyond the largest $g$ shown.

\begin{figure}
\begin{center}
\includegraphics[width=6.8cm]{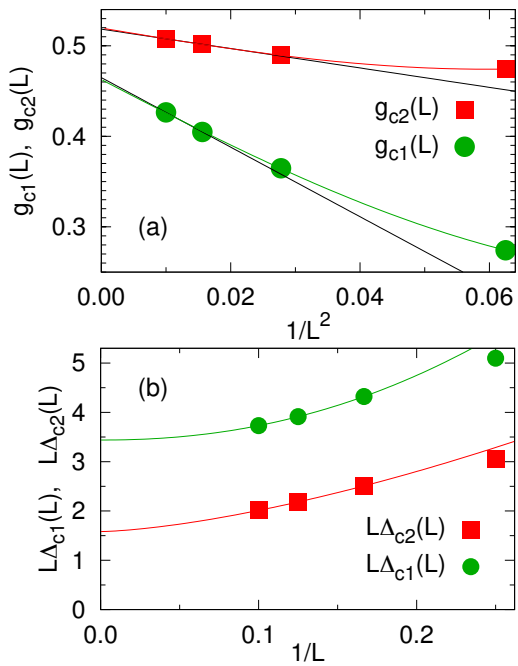}
\vskip-1mm
\caption{(a) The gap-crossing points from Fig.~\ref{gapplot} graphed vs
  $L^{-2}$.  For the singlet-triplet (red squares) and
  singlet-quintuplet (green circles) data sets, the black lines go
  through the $L=8,10$ points, while the colored curves are of the form
  $g_c(L) = g_c(\infty) + aL^{-2}(1+bL^{-\omega})$ with
  $g_{c2}(\infty) \approx 0.519$, $g_{c1}(\infty) \approx 0.463$, and
  $\omega \approx 4$. (b) Size-scaled gaps at the singlet-quintuplet 
  ($\Delta_{c1}$) and singlet-triplet ($\Delta_{c2}$) crossing points 
  along with fits of the form $L\Delta(L)=c+dL^{-\sigma}$, where 
  $\sigma_1 \approx 2$ and $\sigma_2 \approx 1.5$.}
\label{gc}
\end{center}
\vskip-3mm
\end{figure}

As $L$ increases the two sets of crossing points drift toward two different
asymptotic values. For the singlet-triplet crossings, we have considered
different extrapolation procedures with $g_{c2}(L)$, all of which
deliver $g_{c2} \approx 0.52$ when $L \to \infty$.  It is natural to test
whether the finite-size correction to $g_{c2}$ is consistent with the
$L^{-2}$ drift in the frustrated Heisenberg chain
\cite{nomura92,eggert96,sandvik10a}; a behavior also found in the 2D $J$-$Q$
model in Ref.~\onlinecite{suwa_prb94.144416}. In Fig.~\ref{gc}(a) we graph
the data versus $L^{-2}$ along with a line drawn through the $L=8$ and $L=10$
points, as well as a fitted curve including a higher-order correction.
Although we have only four points and there are three free parameters,
it is not guaranteed that the fit should match the data as well as it does. 
With a leading $L^{-1}$ correction the best fit is far from good. Therefore, 
we take the former fit as evidence that the asymptotic drift is at least very 
close to  $L^{-2}$. The fit with the subleading correction in Fig.~\ref{gc}(a) 
gives $g_{c2}=0.519$; a minute change from the straight-line extrapolation. 
Based on the differences between the two extrapolations 
and roughly estimated errors on the individual crossing points (which arise
from the DMRG extrapolations, as discussed in SM \cite{sm}), the final result
is $g_{c2}=0.519 \pm 0.002$.

Plotting the singlet-quintuplet crossing points in the same graph in
Fig.~\ref{gc}(a), the overall behavior is similar to the singlet-triplet
points, but it is clear that they do not drift as far as to $g_{\rm c2}$. We
find that the $L^{-2}$ form applies also here; see the SM \cite{sm} for further 
analysis of the corrections for both $g_{c1}$ and $g_{c2}$. A rough extrapolation 
by a line drawn through the $L=8$ and $L=10$ points gives $g_{\rm c1} \approx 0.465$, 
and when including a correction, of the same form as in the singlet-triplet case, the
extrapolated value moves only slightly down to $g_{\rm c1} \approx 0.463$. Based
on this analysis we conclude that $g_{\rm c1}= 0.463 \pm 0.002$.

In Fig.~\ref{gc}(b) we analyze the crossing gaps, multiplied by $L$ in 
order to make clearly visible the leading behavior and well-behaved corrections.
All gaps close as $L^{-1}$, i.e., the dynamic exponent $z=1$ at both critical 
points. We have also analyzed the gaps in the regime $g_{\rm c1} < g < g_{\rm c2}$ (not shown),
and it appears that the lowest $S=0,1,2$ gaps all scale as $L^{-1}$ throughout.
This phase should therefore be a gapless (algebraic) SL, instead of a $Z_2$ SL with 
nonzero triplet gap for $L\to \infty$ \cite{JiangJ1J2} and singlet gap vanishing 
exponentially (due to topological degeneracy).

The point $g_{c2} \approx 0.52$ is higher than almost all previous results reported 
for the point beyond which the AFM order vanishes, but it is close
to where recent works have suggested a transition from a gapless SL into a
VBS \cite{ShengJ1J2,Imada15}. If there indeed is a gapless SL intervening
between the AFM and the VBS phases and its lowest excitation is a triplet (as
is the case, e.g., in the critical Heisenberg chain), then a singlet-triplet
crossing is indeed expected at the SL--VBS transition, since the triplet is
gapped and the ground state is degenerate in the VBS phase.

To interpret the singlet-quintuplet crossing at $g_{c1} \approx 0.46$, we again 
note that the nature of the low-lying gapless excitations reflect the properties
of the ground state, and a ground state transition can be accompanied by
rearrangements of levels across sectors or within a sector of fixed total
spin. A singlet-quintuplet crossing is indeed present at the transition between 
a critical Heisenberg state (an 1D algebraic SL) and a long-range AFM state
in a spin chain with long-range unfrustrated interactions and either 
unfrustrated \cite{laflorencie} or frustrated \cite{sandvik10a,sandvik10anote}
short-range interactions, as we discuss further in the SM \cite{sm}. 
This analogy, and the fact that $g_{c1}$ is close to where many previous 
works have located the end of the AFM phase (as we also show below and
in SM \cite{sm}), provides compelling evidence for the 
association of the singlet-quintuplet crossing with the AFM--SL transition.
Furthermore, the $S=2$ quantum rotor state in the AFM state has gap
$\propto L^{-2}$, while at $g_{c1}$ it scales as $L^{-1}$ according
to Fig.~\ref{gc}. Thus, at this point (and for higher $g$) the level spectrum
is incompatible with AFM order.

\begin{figure}
\begin{center}
\includegraphics[width=7.5cm]{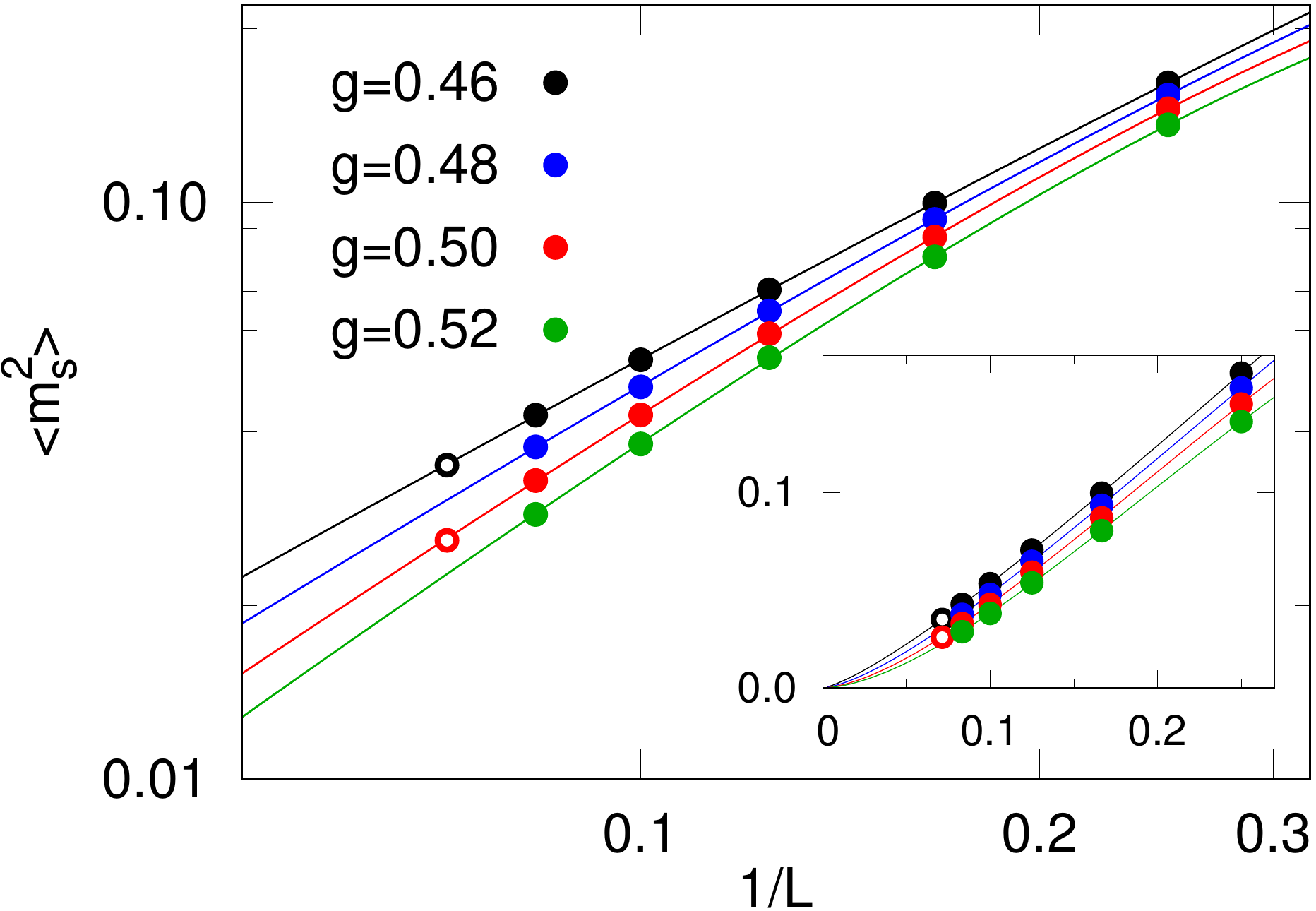}
\vskip-1mm
\caption{Log-log plot of $\langle m_s^2\rangle$ vs $L^{-1}$. The curves are
  of the form $\langle m_s^2\rangle = bL^{-\alpha}(1-cL^{-\omega})$ with
  $\omega=0.5$. The leading exponent, with errors estimated by changing
  $\omega$ within its range of good fits, are $\alpha=1.35 \pm 0.05 $
  ($g=0.46$), $1.53 \pm 0.08$ ($g=0.48$), $1.69 \pm 0.10$ ($g=0.50$), and
  $1.78 \pm 0.12$ ($g=0.52$). The inset shows the same data on a linear scale.
  The $L=14$ data (open circles) are from Ref.~\onlinecite{ShengJ1J2}.} 
\label{mag}
\end{center}
\vskip-2mm
\end{figure}

We also computed the squared AFM order parameter (sublattice magnetization
per spin) $\langle m^2_s\rangle$ in the putative SL phase, with ${\bf m}_s$
defined on the central $L\times L$ part of the $2L\times L$ system (here with
$L$ up to $12$). Since we mainly focused on the excited energies, we did
not push the ground state $\langle m^2_s\rangle$ calculations to as large
$L$ as in some past works \cite{JiangJ1J2,ShengJ1J2}. To complement
our own data, we therefore also use $L=14$ results from
Ref.~\onlinecite{ShengJ1J2}. In cases where we have data for the same
parameter values, our results agree to within $0.2\%$. We fit the data 
to power laws with a correction; $\langle m_s^2\rangle = bL^{-\alpha}(1-cL^{-\omega})$,
where acceptable values of $\omega$ span the range $\omega \approx 0.2 \sim 1.5$
and the exponent $\alpha$ changes somewhat when varying $\omega$. In Fig.~\ref{mag} 
we show examples of fits with $\omega=0.5$. We find that $\alpha$ increases with $g$, 
from $\alpha \approx 1.3$ at $g=0.46$ to $\alpha \approx 1.8$ at $g=0.52$. We have
also tried to fix $\alpha$ to a common value for all $g$, but this does not
produce good fits. We therefore agree with previous claims
\cite{ShengJ1J2,Imada15} that the exponent depends on $g$. At $g=0.5$, 
our result $\alpha \approx 1.7 \pm 0.1$ is larger than the
value $1.44$ reported in Ref.~\onlinecite{ShengJ1J2}, with the difference
explained by the correction used here. The result agrees well with
$\alpha =1.53 \pm 0.09$ from variational Monte Carlo calculations
\cite{Imada15}, and a similar value was also reported with a projected 
entangled pair state ansatz \cite{mambrini17}. In the SM \cite{sm} 
we provide further analysis showing that the AFM order vanishes at the
extrapolated level crossing point $g_{c1} \approx 0.46$. 

{\it Discussion.---}Our level-crossing analysis in combination with results
for the sublattice magnetization show consistently that the AFM phase ends 
at $g_{\rm c1} \approx 0.46$ and a gapless SL phase exists between this value
and $g_{\rm c2} \approx 0.52$.  In the level crossing approach the
finite-size transition points are sharply defined and the convergence with
system size is rapid, with corrections vanishing as $L^{-2}$ (or possibly 
$L^{-a}$ with $a\approx 2$). Our results in Fig.~\ref{gc}(a) leave little doubt 
that the singlet-quintuplet and singlet-triplet crossings converge to different
points, while we would expect convergence to the same point if there is no 
SL between the AFM and VBS phases, as we demonstrate explicitly in the
SM~\cite{sm} in the case of the $J$-$Q$ model. The behavior of the spin
correlations and the gaps imply a gapless SL with power-law decaying spin
correlations. In the region $0.52 < g < 0.62$, between the SL and the
stripe-AFM, our calculations of excited states reveal many low-lying
singlets, and we have been able to map them \cite{tobeappear} onto 
the expected quasi-degenerate levels expected for a columnar 
\cite{Imada15} VBS state.

The AFM--SL and SL--VBS phase boundaries are in rough agreement with two
recent works discussing a gapless SL phase followed by a VBS
\cite{ShengJ1J2,Imada15}, and the lower boundary agrees well with a
Lanczos-improved variational Monte Carlo calculation \cite{HuJ1J2}.  Many
other past studies have located the end of the AFM order close to the same
value. A recent exception is an infinite-size tensor calculation
\cite{HaghshenasJ1J2} where the AFM order ends close to our $g_{c2}$ point.
However, the infinite-size approach is not unbiased but depends on details of
how the environment tensors are constructed. The DMRG calculations, here and
in Ref.~\onlinecite{ShengJ1J2}, are unbiased for finite size if the
convergence is checked carefully, and completely exclude AFM order beyond our
$g_{c1}$ value.

As far as we are aware, the critical singlet-quintuplet crossing found here 
(and the singlet-singlet crossing in the SM \cite{sm}) has not previously 
been discussed in the 2D context. This level crossing has been considered in 
1D \cite{sandvik10a,sandvik10anote}, and in the SM \cite{sm} we present additional 
evidence of its association with the AFM--SL transition. The physical 
origin of the level crossing deserves further study. The detailed information 
we have obtained on the evolution of the low-energy levels in 2D should be 
useful for discriminating between different field theoretical descriptions 
of the phase transitions and the SL phase.

We expect that level crossings are common at 2D quantum phase transitions, as they
are in 1D. Our work suggests that the best way to use 2D DMRG in studies of quantum 
criticality is to first look for and analyze level crossings to extract critical points, 
and then study order parameters (conventional or topological) at this point and in the
phases. In principle the DMRG procedures that we have employed here can also
be extended to more detailed level-spectroscopy studies
\cite{suwa_prb94.144416,schuler16}.

\begin{acknowledgments} {\it Acknowledgments.---}We would like to thank
  F.~Becca, S.~Capponi, M.~Imada, D.~Poilblanc, S.~Sachdev, J.-Z.~Zhao, and
  Z.-Y.~Zhu, for helpful discussions. We are grateful to S. Gong and D. Sheng
  for providing their numerical results from
  Ref.~\onlinecite{ShengJ1J2}. L.W. is supported by the National Key Research
  and Development program of China (Grant No.~2016YFA0300600), the National
  Natural Science Foundation of China (Grant No.~NSFC-11734002 and
  No.~NSFC-11474016), the National Thousand Young Talents Program of China,
  and the NSAF Program of China (Grant No.~U1530401). She thanks Boston
  University's Condensed Matter Theory Visitors program for travel support.
  A.W.S. was supported by the
  NSF under grants No.~DMR-1410126 and DMR-1710170, and by a Simons Investigator Grant. 
  He would also like to thank the Beijing Computational Science Research Center (CSRC) for visitor support. 
 The calculations were partially carried out under a Tianhe-2JK computing award at the CSRC.
  
\end{acknowledgments}

\newpage

\setcounter{page}{1}
\setcounter{equation}{0}
\setcounter{figure}{0}
\renewcommand{\theequation}{S\arabic{equation}}
\renewcommand{\thefigure}{S\arabic{figure}}

\section*{Supplemental Material}
\vskip-2mm

\subsection*{Critical level crossings in the square-lattice spin-1/2 J1-J2 Heisenberg antiferromagnet}
\vskip-2mm

\centerline{Ling Wang and Anders W. Sandvik}
\vskip5mm

We have argued that the AFM--SL transition in the 2D $J_1$-$J_2$ Heisenberg model is associated
with a level crossing between the lowest singlet excitation and the first quintuplet
($S=2$), while the singlet-triplet crossing is associated with the SL--VBS transition.
We here provide further supporting evidence for this scenario.

In Sec.~I, we first
illustrate our stringent DMRG convergence checks and extrapolations of the low-energy
levels.
In Sec.~II, we contrast the findings for the $J_1$-$J_2$ model with results for
the $J$-$Q$ model, where it is known that no SL phase intervenes between the AFM and VBS
states. Accordingly, we show that the singlet-triplet and singlet-quintuplet crossing points
flow with increasing system size to the same critical point (a deconfined quantum-critical
point). We also investigate the critical scaling of the sublattice magnetization of the
$J$-$Q$ model on the cylinders and compare with the $J_1$-$J_2$ model. In Sec.~III we
present further tests of the scaling behavior of the level crossing points and the
sublattice magnetization of the $J_1$-$J_2$ model. The singlet-quintuplet crossing in the 
2D $J_1$-$J_2$ model is analogous to a crossing point previously found in a spin chain with 
long-range interactions at its transition from a critical SL phase to an AFM phase 
\cite{sandvik10a,sandvik10anote,laflorencie}. In Sec.~IV we provide further results for the 1D model, 
using the excited-level DMRG method to go to larger system sizes than in the past Lanczos
calculations. In the 2D $J_1$-$J_2$ model, in addition to the singlet-quintuplet
crossing at the AFM--SL transition, we also find a crossing between the two lowest
singlet excitations, and in Sec.~V we present the numerical results and analysis
of this level crossing.

\subsection*{I. DMRG convergence procedures}

\noindent
In each DMRG calculation bounded by $m$ Schmidt states, we start from a previously
converged MPS with a smaller $m$ and perform a number of DMRG sweeps until the energy
converges sufficiently. The convergence criterion for an $m$-bounded MPS is that the total
energy difference (i.e., not the difference in the average energy per site) between 
two successive full sweeps is less than $2\times 10^{-6}$, which we have confirmed to be sufficient
by comparing with calculations done with less stringent criteria. We then check
the convergence of the energies as a function of the discarded weight $\epsilon$
(which depends on $m$, with $\epsilon \to 0$ as $m \to \infty$) defined in the
standard way in DMRG calculations as the sum of discarded eigenvalues of the
reduced density matrix.

\begin{figure}[t]
\begin{center}
\includegraphics[width=6.5cm]{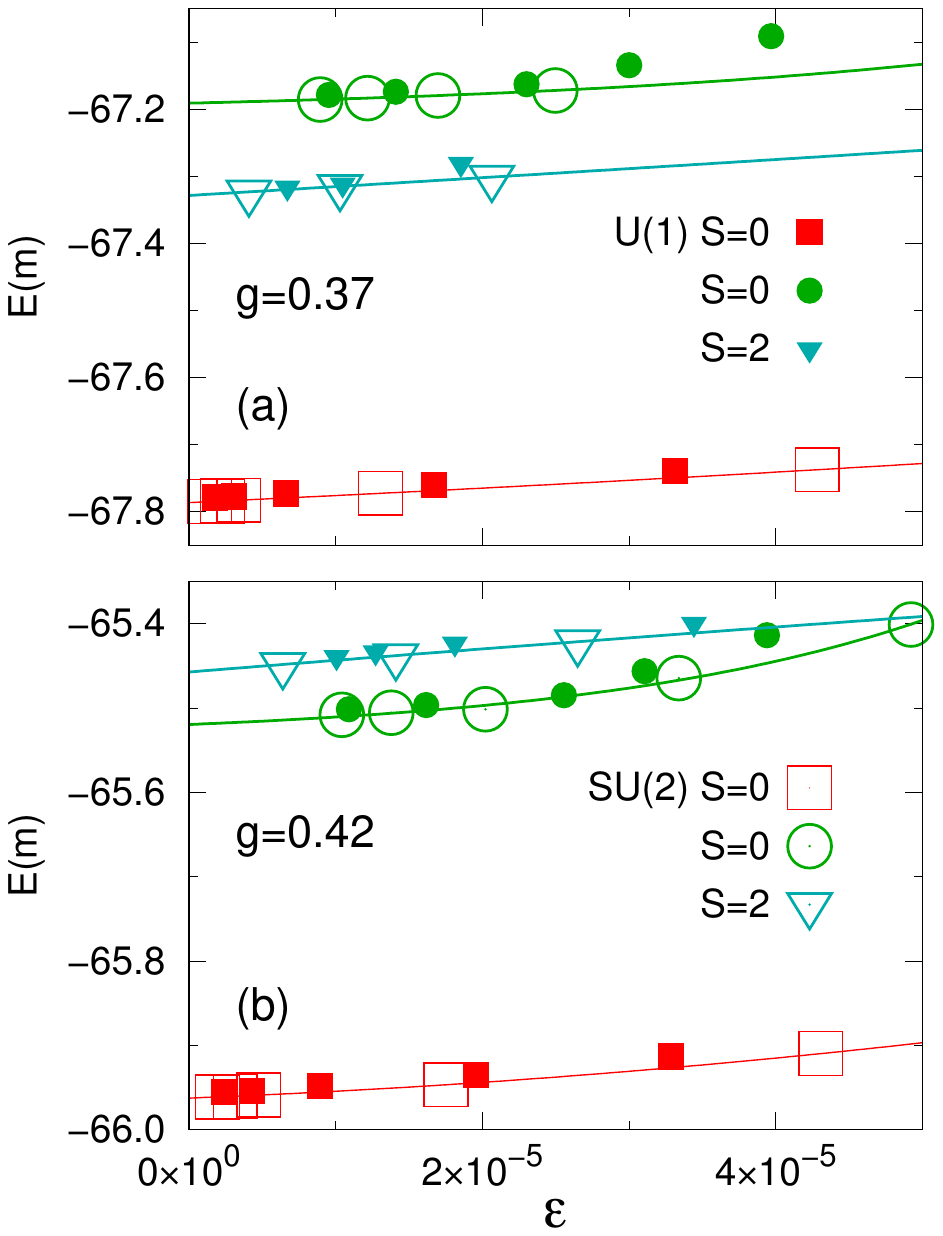}
\vskip-1mm
\caption{DMRG energies for $L=8$ graphed vs the
  discarded weight $\epsilon$. The largest number of Schmidt states was
  $m=2000$ and  $m=4000$ for SU(2) and U(1) symmetry, respectively. 
  The curves are fits to constants plus exponential corrections. In both panels,
  open symbols represent SU(2) energies and the corresponding filled symbols
  represent U(1) energies; squares for the ground state, circles for the lowest singlet
  excitation, and triangles for the lowest quintuplet. }
  \label{varextrapolation}
\end{center}
\vskip-2mm
\end{figure}

In Fig.~\ref{varextrapolation} we show the convergence of the first two $S=0$
energies and the first $S=2$ level for an $L=8$ system at two $g$ values close
to $g_{c1}$ (the AFM--SL transition), using $m$ up to $4000$ in calculations
with U(1) symmetry and $m$ up to $2000$ with SU(2) symmetry. In our analysis
of the AFM-SL transition we used a singlet-quintuplet crossing in the main paper,
and in Sec.~V we will also investigate the excited singlet-singlet crossing.
 With SU(2) symmetry implemented, the lowest state in calculations with $S=0$
fixed is the ground state, and we make sure to converge two additional states in this spin sector.
At the AFM--SL transition, we further carry out calculations with $S=2$ for the lowest quintuplet.
With only U(1) symmetry, the lowest state in the $S^z=0$ sector is the ground state, while
the lowest state with $S^z=2$ is also the lowest excitation with $S=2$. To compute
the two lowest singlet excitations close to the AFM--SL transition for $L=8$, one has
to go to 6th and 7th excitations in the $S^z=0$ sector in the case of $L=8$. In Fig.~\ref{varextrapolation}
the SU(2) DMRG eigenvalues nevertheless coincide very well with the corresponding U(1)
energies in all cases when $\epsilon$ is small. All the states show exponentially fast 
convergence when $\epsilon \to 0$, and we can obtain stable extrapolated energies.

\begin{figure}[t]
\begin{center}
\includegraphics[width=6.5cm]{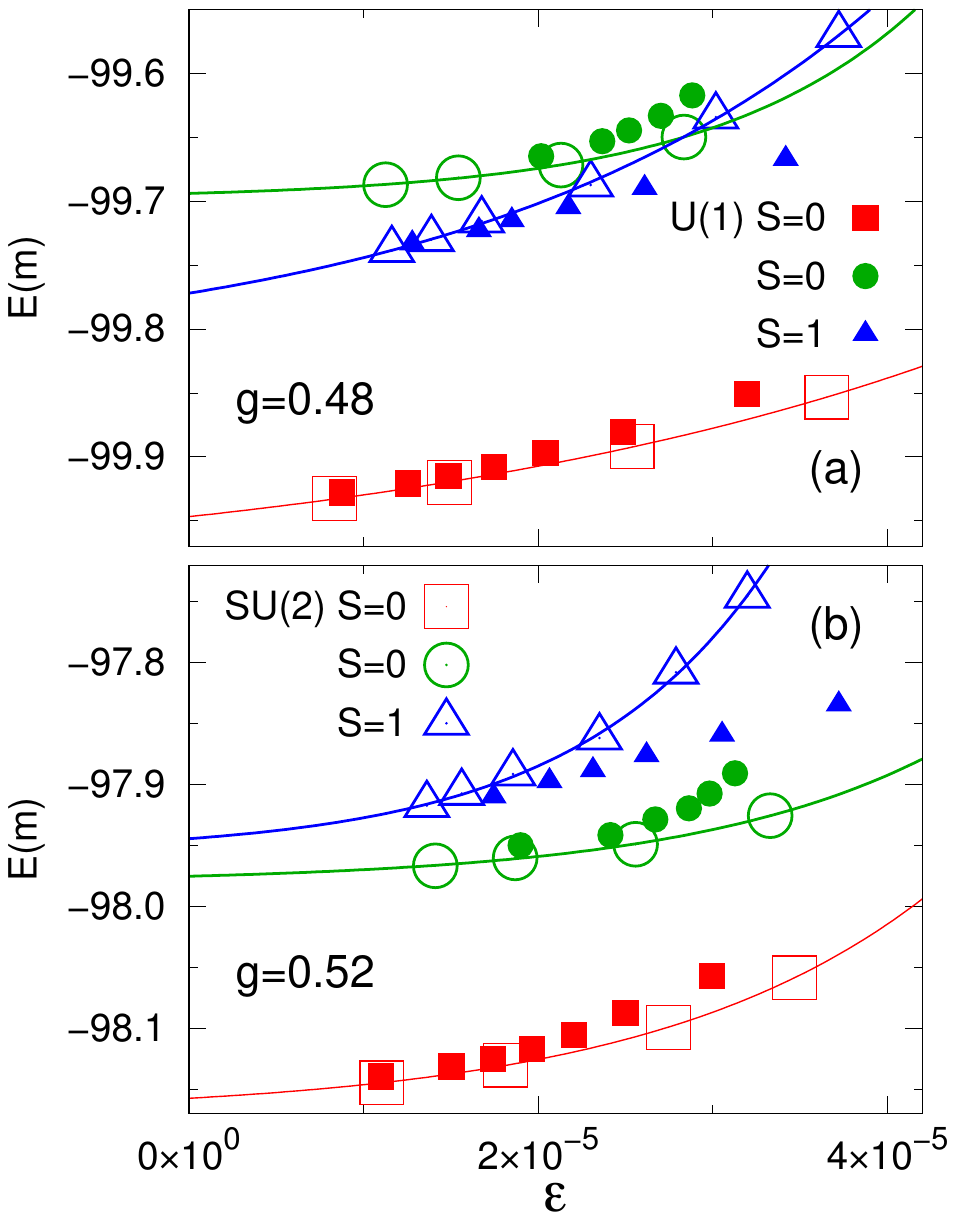}
\vskip-1mm
\caption{DMRG energies for $L=10$ graphed vs the
  discarded weight $\epsilon$. The largest number of Schmidt states was
  $m=5000$ and  $m=12000$ for SU(2) and U(1) symmetry, respectively. 
 The curves are fits to constants plus
  exponential corrections. In both panels, open symbols represent SU(2)
  energies and the corresponding filled symbols represent U(1) energies:
  squares for the ground state, circles for the lowest singlet
  excitation, and triangles for the lowest triplet.}
\label{enrextrapolation}
\end{center}
\vskip-2mm
\end{figure}

For $L=10$, we show the energy convergence at two $g$ values close to
$g_{c2}$ (the SL--VBS transition) in Fig.~\ref{enrextrapolation}, using $m$
up to $12000$ with U(1) symmetry and $m$ up to $5000$ with SU(2)
symmetry. The SL--VBS phase transition is detected as the level crossing
between the lower singlet and the lowest triplet. With SU(2) symmetry, the
lowest state in the $S=0$ sector is the ground state and the lowest triplet
is the ground state in the $S=1$ sector.  To obtain the lowest singlet
excitation used in our analysis in the nonmagnetic state, we target the
second $S=0$ state near $g_{c2}$. With U(1) symmetry, the lowest $S^z=0$
state is the ground state, while the lowest state in the $S^z=1$ sector is
the lowest triplet excitation.  To compute the first excited singlet for
$g_{c1} < g \alt g_{c2}$, we need to target the third level with $S^z=0$
(since one of the triplet states also has $S^z=0$ and is lower in energy than
the targeted singlet) but only need the first excitation when $g>g_{c2}$
(since the triplet is higher there). As seen in Fig.~\ref{enrextrapolation},
for small $\epsilon$ the SU(2) and U(1) energies again coincide very well.

We regard the essentially perfect agreement between the SU(2) and U(1) calculations
for large $m$ (in the $L=8$ and $10$ demonstrations above as well as in other cases studied) as
evidence for sufficient convergence in both cases. We have estimated the remaining
small systematical errors by comparing the U(1) and SU(2) extrapolations in detail
and by varying the functional form used in the extrapolations.

\subsection*{II. Critical level crossings and order parameter of the J-Q model on a cylinder}

In Ref.~\onlinecite{suwa_prb94.144416}, the critical level crossings of the lowest
singlet and triplet excitation in $J$-$Q$ model were studied using quantum Monte
Carlo (QMC) simulations of $L\times L$ lattices with fully periodic (torus)
boundaries. The decay rates of the
spin-spin and dimer-dimer correlation functions in imaginary time were used
to extract the gaps in the triplet and singlet channels, respectively. It was
found that the finite size level crossing points $g_c(L)$ approach a value
$g_c$ that is fully consistent with the AFM--VBS quantum critical point previously
extracted by finite-size scaling of the order parameters. The scaling correction
was found to be $g_c(L) - g_c \propto L^{-2}$. The level crossing in this case
is expected, given the known behaviors of the lowest singlet and triplet
in the AFM and VBS states.

In the main text, we concluded that the $J_1$-$J_2$ model hosts an SL phase between
the AFM and VBS states and that the AFM-SL transition is associated with a crossing
between $S=0$ and $S=2$ excitations. It is then interesting to look for and investigate
singlet-quintuplet level crossings also in the $J$-$Q$ model, as a test that a
second, spurious  critical point is not found in this case. In addition, it is also
useful to study the singlet-triplet crossings with the same DMRG method that we
have used for the $J_1$-$J_2$ model, and with the same cylindrical lattices, to
check that we can correctly reproduce the AFM-VBS transition point even in
this geometry and with the much more limited system sizes than in the QMC
calculations. A related question is whether the change of lattice geometry
will affect the power-law scaling behavior of the finite-size size crossing
points $g_c(L)$.

\begin{figure}
\begin{center}
\includegraphics[width=6.5cm]{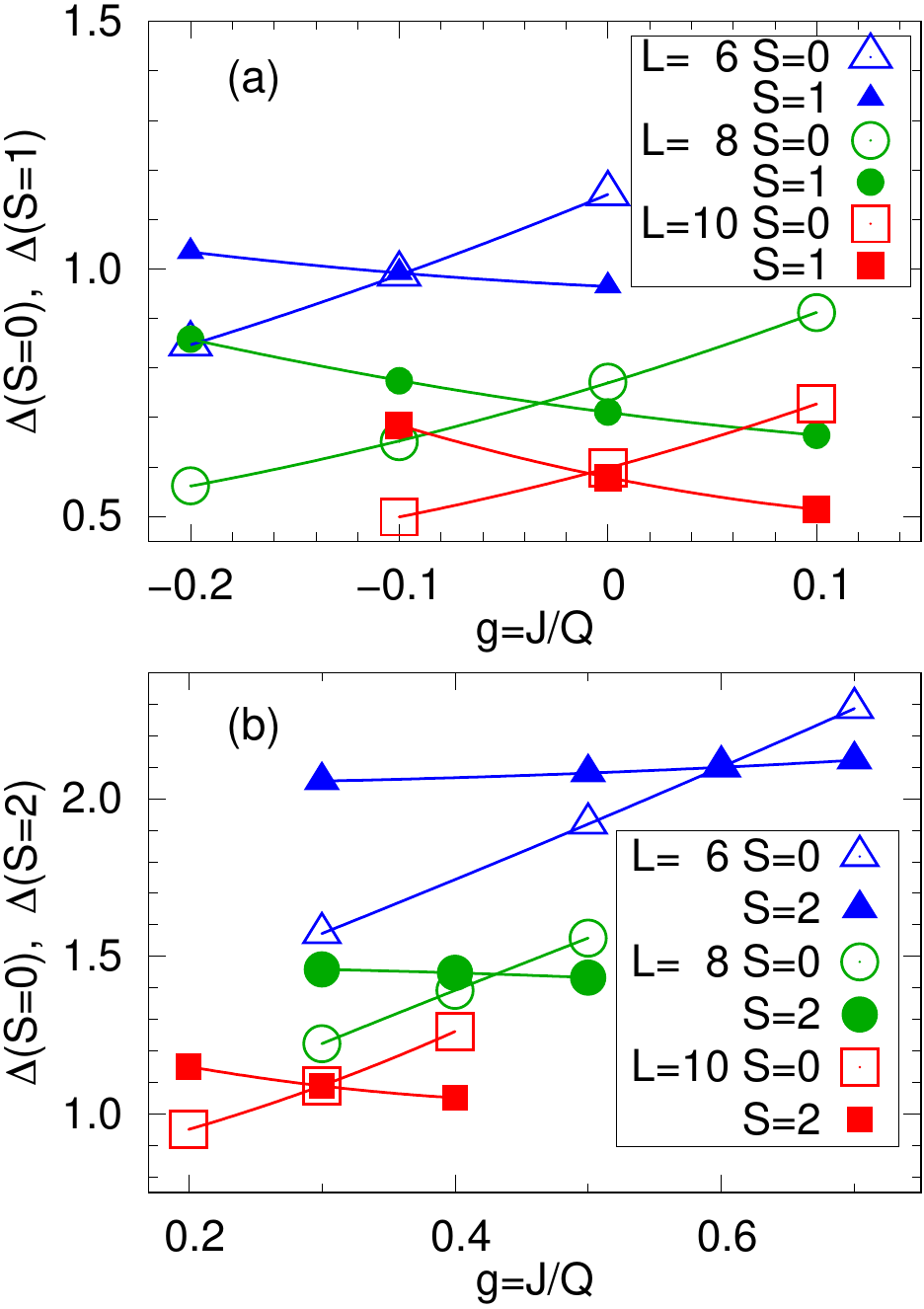}
\vskip-1mm
\caption{Gaps vs $g=J/Q$ of the $J$-$Q$ model on $2L\times L$ cylinders. (a) The lowest singlet and
  triplet gaps. (b) The lowest singlet and quintuplet gaps. Both crossing
  points $g_{c1}(L)$ (singlet-quintuplet) and $g_{c2}(L)$ (singlet-triplet)
  are extracted using second-order polynomial fits (the curves shown).}
\label{jqlevelcross}
\end{center}
\vskip-2mm
\end{figure}

We study the lowest singlet-triplet and singlet-quintuplet gap
crossings in the standard $J$-$Q$ model~\cite{DQCP3}, using the DMRG
method with U(1) symmetry on $2L\times L$ cylinders with $L=4,6,8,10$.
Before presenting the DMRG results, we recall some of the well studied
ground state properties of the model from previous QMC simulation in both
the torus and cylinder geometries ~\cite{DQCP3,sandvik12}. At
$Q=0$, the $J$-$Q$ model reduces to the standard 2D Heisenberg model with
AFM order, while at $J=0$ the ground state is a columnar VBS with four-fold
degeneracy on a torus. When tuning the coupling ratio  $g\equiv J/Q$ from
$+\infty$ to $0$, the system goes through a deconfined quantum phase transition
from the AFM phase to the columnar VBS phase at $g_c\approx 0.045$, where the
lowest singlet and triplet gaps cross each other when $L \to \infty$ as mentioned
above. In addition, it is known that the ground state of the $J$-$Q$ model on
$2L\times L$ cylinders in the VBS phase is a non-degenerate columnar VBS state
with $x$-oriented dimers. In our DMRG calculations presented below, we resolve
that, in the VBS phase, the ground state has momentum $k_y=0$, and above it
there is a singlet excited state with momentum $k_y=\pi$. The $k_y=\pi$
singlet, which is related to the open $x$-direction boundary condition, lies
below the first triplet excitation and remains with a non-vanishing gap to
the unique ground state in the thermodynamic limit.

\begin{figure}
\begin{center}
\includegraphics[width=6.5cm]{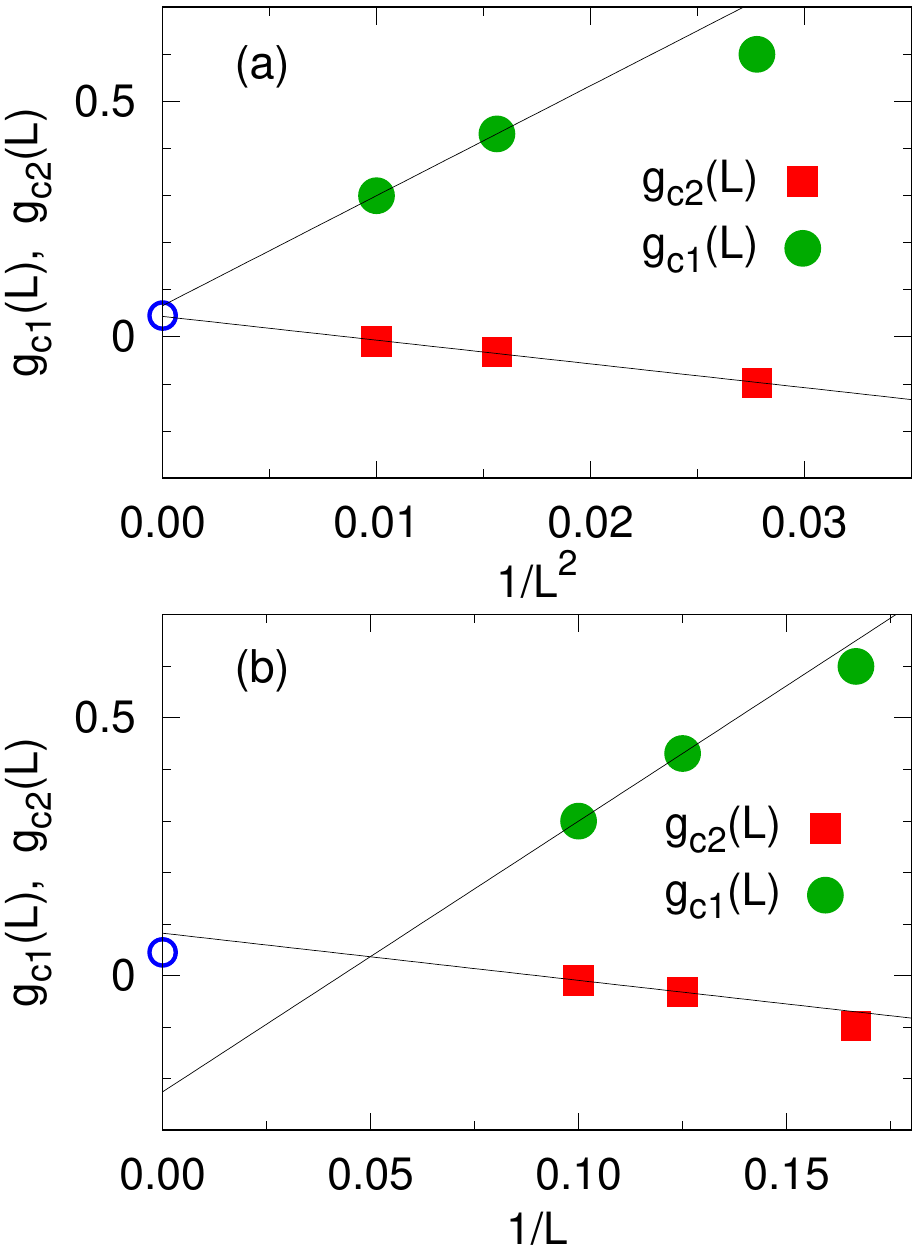}
\vskip-1mm
\caption{Size dependence of the singlet-triplet (red squares) and singlet-quintuplet (green circles) gap 
crossing points extracted from the data in Fig.~\ref{jqlevelcross}. (a) The crossing points graphed vs $L^{-2}$. 
The lines are drawn through the $L=8$ and $L=10$ points and give the extrapolated crossing values $g_{c2}(\infty) 
\approx 0.043$ and $g_{c1}(\infty) \approx 0.066$. Both these fall close to the known AFM--VBS transition of 
the $J$-$Q$ model; $g_c = 0.045$ (indicated by the blue circle). (b) The crossing points graphed vs $L^{-1}$, 
with lines drawn through the $L = 8$ and $L=10$ points. Here the extrapolated crossing points 
deviate significantly from the critical point.}
\label{jq_gc}
\end{center}
\vskip-2mm
\end{figure}

Figure \ref{jqlevelcross} shows the gaps versus $g$ on $2L\times L$ cylinders
with $L=6,8,10$, with singlets and triplets analyzed in (a), and the singlets and
quintuplets in (b). We fit second order polynomials to the data and interpolate for 
the crossing points. As $L$ increases, the singlet-triplet crossing points $g_{c2}(L)$ 
drift toward $g_c$ from the left, while the singlet-quintuplet crossing points $g_{c1}(L)$ 
drift toward $g_c$ from the right. 

It is again natural to check whether the finite-size corrections to the
crossing points $g_c$ is consistent with the same form, $L^{-2}$, as in the
model on a torus. Fig.~\ref{jq_gc}(a) shows $g_{c1}(L)$ and $g_{c2}(L)$
versus $L^{-2}$ along with a line drawn through the $L=8,10$ points.
These simple extrapolations give $g_{c2}=0.043$ (singlet-triplet) and $g_{c1}=0.066$
(singlet-quintuplet). Considering the small systems and the extrapolation
without any corrections, these results are both in reasonable agreement with
the known critical point, $g_c \approx 0.045$. The results also support
leading $L^{-2}$ corrections for the cylindrical lattices and lend further
credence to our use of this form of the corrections in the $J_1$-$J_2$ model.
In contrast, if we assume that the crossing points drift as $L^{-1}$, as shown 
in Fig.~\ref{jq_gc}(b), the extrapolated points $g_{c2}$ and $g_{c1}$
are very different and disagree with the known critical coupling.

\begin{figure}
\begin{center}
\includegraphics[width=6.5cm]{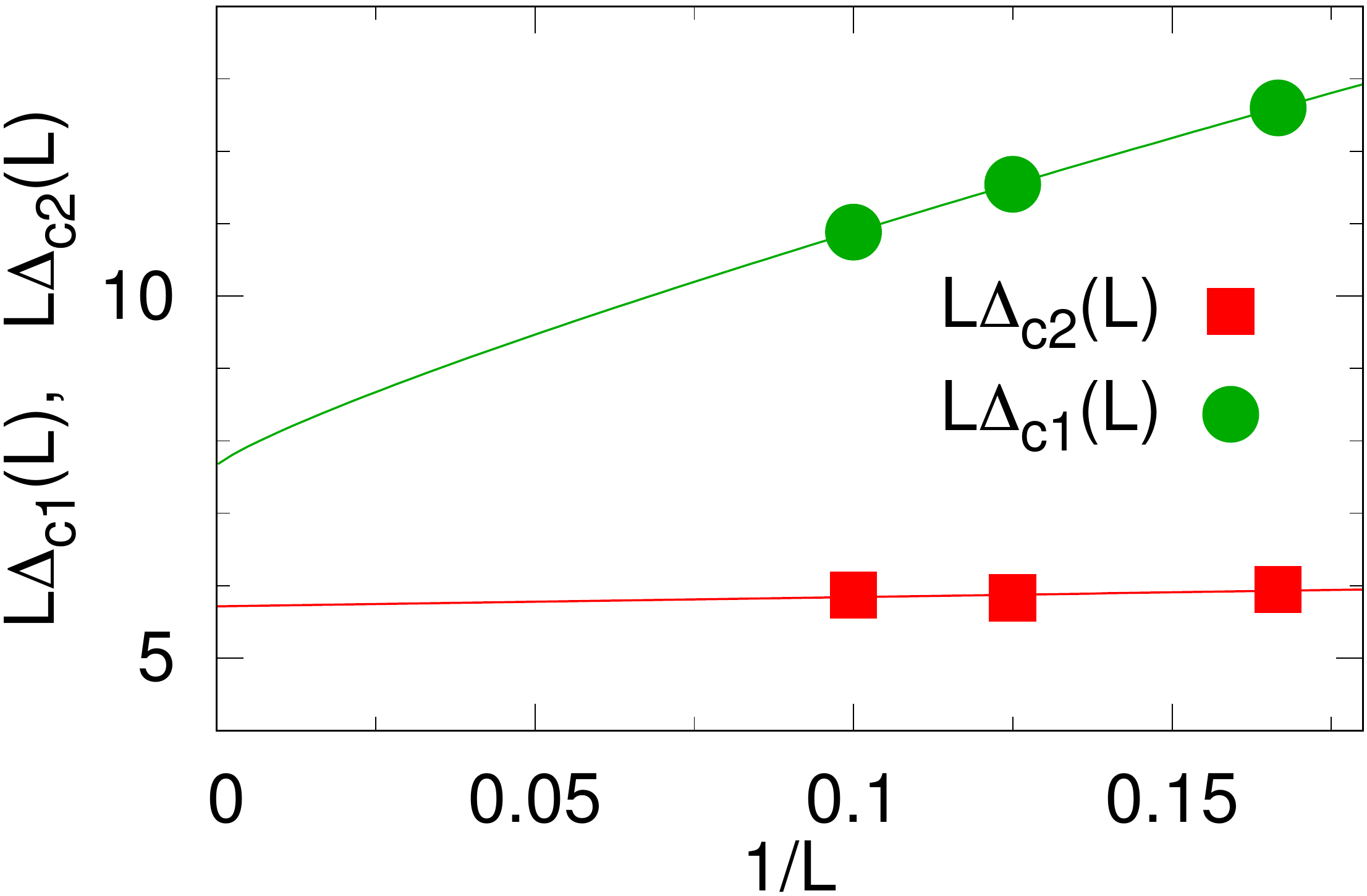}
\vskip-1mm
\caption{Size-scaled gaps of the $J$-$Q$ model at the size dependent 
singlet-quintuplet ($\Delta_{c1}$) and singlet-triplet ($\Delta_{c2}$) crossing 
points from Fig.~\ref{jqlevelcross}. The curves are fits with power-law corrections 
to the infinite-size values.}
\label{jq_gap}
\end{center}
\vskip-2mm
\end{figure}

We analyze the gaps $\Delta_{c1}(L)$ and $\Delta_{c2}(L)$ of the $J$-$Q$ model  
at the $L$-dependent crossing points in Fig.~\ref{jq_gap}. We have multiplied the
gaps by $L$ and graph the results versus $L^{-1}$. We see clear signs of convergence 
to constants, confirming that the gaps close as $L^{-1}$ at the critical
point, as expected since the dynamic critical exponent is $z=1$.

\begin{figure}
\begin{center}
\includegraphics[width=7cm]{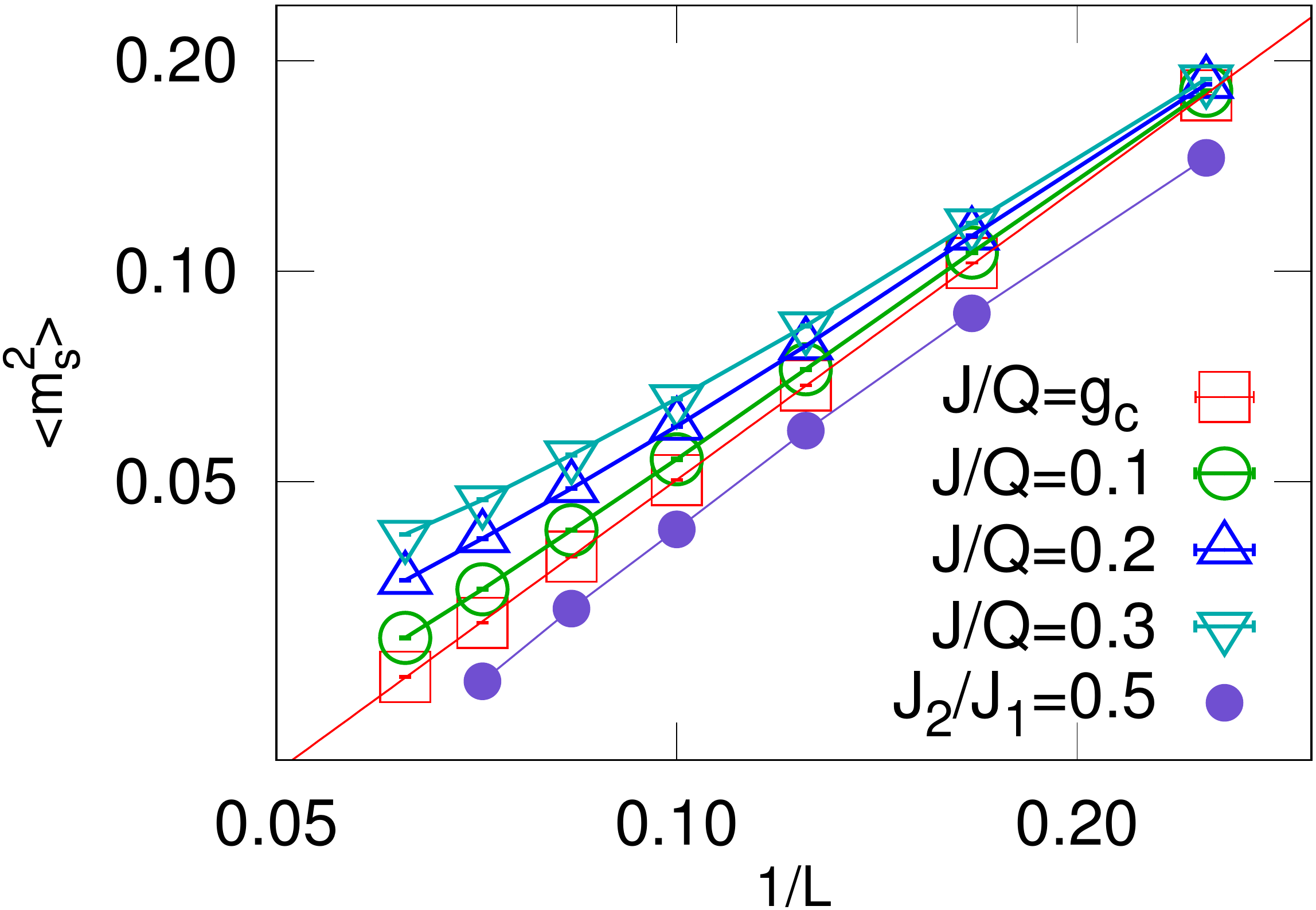}
\vskip-1mm
\caption{Log-log plot of $\langle m_s^2\rangle$ of the $J$-$Q$ model vs $L^{-1}$.
The results were computed in the central $L\times L$ square of $2L\times L$ cylinders. The 
red line is a linear fit (power-law scaling) at $g=0.0447 \approx g_c$. Filled circles are 
for the same quantity in the $J_1$-$J_2$ model at $J_2/J_1=0.5$ (our own DMRG results for 
$L \le 12$ and the $L=14$ point from from Ref.~\onlinecite{ShengJ1J2}).}
\label{jqstgm}
\end{center}
\vskip-2mm
\end{figure}

Next, we consider the AFM order parameter of the $J$-$Q$ model, the squared staggered
magnetization. Fig.~\ref{jqstgm} shows $\langle m_s^2\rangle$ computed in
the center $L\times L$ section of $2L\times L$ cylinders at various coupling ratios
$J/Q$. The results are graphed versus $L^{-1}$ on log-log scales, along with the 
results for the same quantity (defined in the same way on the central parts of the
cylinders) for $J_1$-$J_2$ model at $J_2/J_1=0.5$. Here the results for the $J$-$Q$
model are obtained from QMC simulation (with the same cylindrical boundary conditions that 
we use in the DMRG calculations), in order to reach the same system sizes as for the $J_1$-$J_2$ 
model. The red line on the log-log plot corresponds to a power-law form of $\langle m^2_s\rangle$ 
at $g_c$. Away from $g_c$, inside the AFM phase, we observe that $\langle m^2_s\rangle $
curves upward for the larger sizes relative to the critical power law behavior, as expected 
when the order parameter scales to a non-zero value. This is in
contrast to the behavior in the case of the  $J_1$-$J_2$ model at $J_2/J_1=0.5$,
where $\langle m^2_s\rangle$ decays almost in the same way as in the critical
$J$-$Q$ model, though on close examination one can see a clear downward trend with
increasing size. It therefore appears very unlikely that a non-zero
value would survive in the $J_1$-$J_2$ model when $L \to \infty$; thus the results 
lend further support to the SL scenario.

\subsection*{III. Additional tests of scaling in the $J_1$-$J_2$ model}

\begin{figure}
\begin{center}
\includegraphics[width=6.5cm]{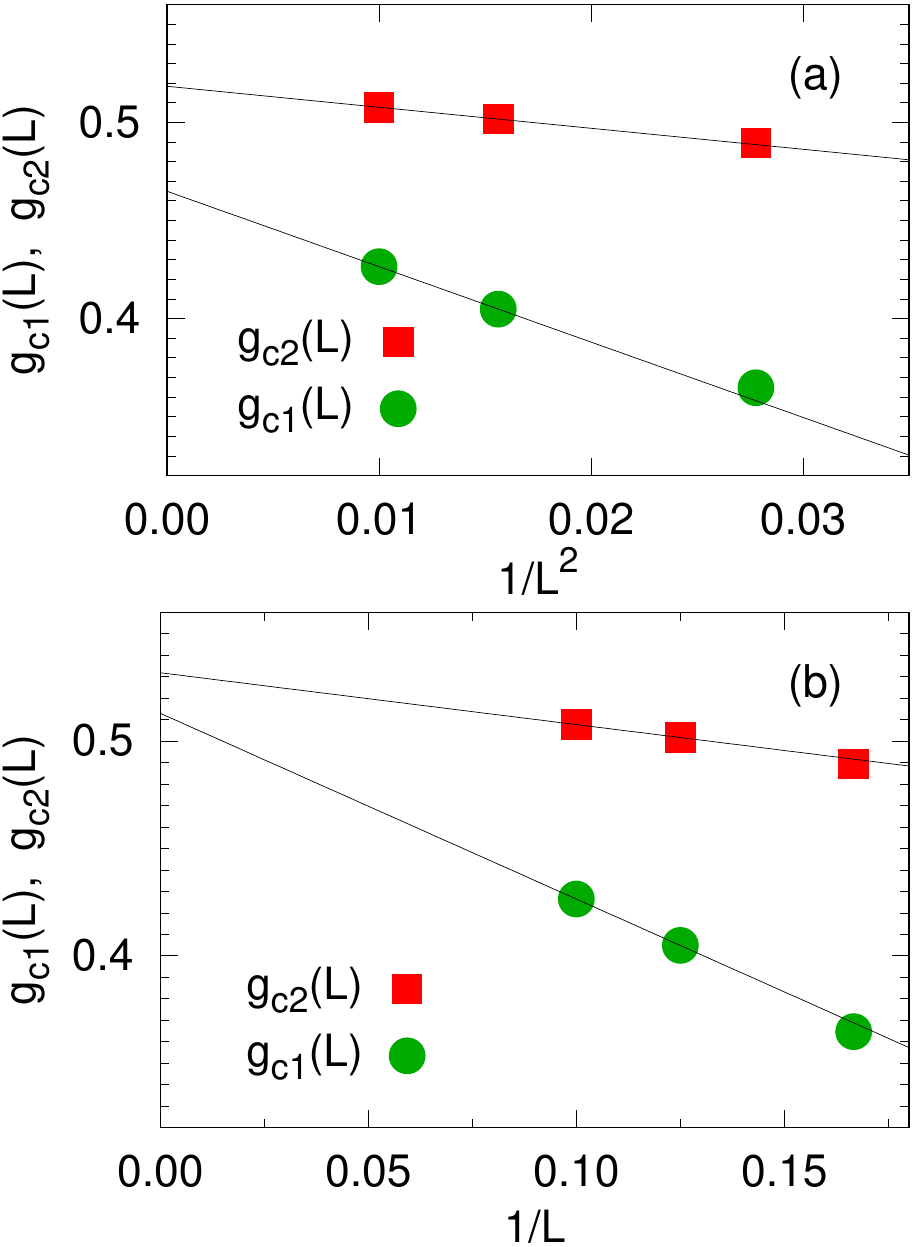}
\vskip-1mm
\caption{(a) The crossing points of the $J_1$-$J_2$ model on lattice sizes $L=6,8$,
and $10$ (from Fig.~2 in the main paper) graphed vs $L^{-2}$ with leading-order extrapolations 
with the $L=8$ and $L=10$ data. (b) The same data graphed vs $L^{-1}$ together with 
leading-order linear fits.}
\label{j1j2_gc}
\end{center}
\vskip-2mm
\end{figure}

In the main text we showed that leading $L^{-2}$ corrections also 
describe well the drifts of crossing points in the case of the $J_1$-$J_2$
model. In Fig.~\ref{j1j2_gc}(a) we again show the results for $L=6,8,10$
(leaving out $L=4$ for clarity) graphed against $L^{-2}$ together with a simple
fit based on just the two largest system sizes. Figure \ref{j1j2_gc}(b) shows 
the same data plotted versus $L^{-1}$, again along with extrapolations using only the 
two largest system sizes. Since the overall size dependence of the singlet-triplet 
crossing is weak, its extrapolation only changes marginally from the one based 
on the $L^{-2}$ form. An extrapolation with a higher-order correction (not shown in the
figure) shifts the value down even closer to the previous estimate. The $L^{-1}$ extrapolated 
singlet-quintuplet 
point is significantly higher then previously, but looking at the trend including  
the smaller sizes makes it clear that higher-order fits here will also reduce the 
extrapolated value. As mentioned in the main text, such higher-order fits do not 
match the data as well as in the case of leading $L^{-2}$ corrections. 

These results with different fitting forms lend support to the existence of a gap
between the extrapolated $g_{c1}$ and $g_{c2}$ values in the $J_1$-$J_2$ and the absence
of such a gap in the $J$-$Q$ model. In the main text we have argued that 
$g_{c1} \not= g_{c2}$ reflects the presence of an SL phase intervening between
the AFM and VBS phases in the $J_1$-$J_2$ model, while $g_{c1} = g_{c2}$ reflects
the known deconfined quantum-critical AFM--VBS point in the $J$-$Q$ model. The
well established $L^{-2}$ scaling in the latter case, from large-scale QMC
simulations \cite{suwa_prb94.144416} as well as the results in Sec.~II above, allow 
us to make a further argument against the deconfined quantum-criticality scenario in 
the $J_1$-$J_2$ model: If the two models both host critical AFM--VBS points, based on 
the deconfined universality class, they should also both
exhibit leading $L^{-2}$ drifts of the crossing points and common extrapolated crossing 
points $g_{c1}=g_{c2}$. However, the results shown in Fig.~\ref{j1j2_gc}(a) and Fig.~3
in the main paper are inconsistent with a common crossing point, unless the system 
sizes we have access to here are not yet in the asymptotic regime where scaling
with small corrections is applicable. While we cannot in principle exclude that a 
cross-over to a single point, a direct AFM--VBS transition, occurs on some larger length scale, 
we see no a priori physical reason for such large finite-size effects (given their
absence in the $J$-$Q$ model) and find this scenario unlikely. Thus, based on all the present 
evidence we conclude that the deconfined critical point most likely is expanded into a
stable nonmagnetic phase in the $J_1$-$J_2$ model.

\begin{figure}
\begin{center}
\includegraphics[width=7cm]{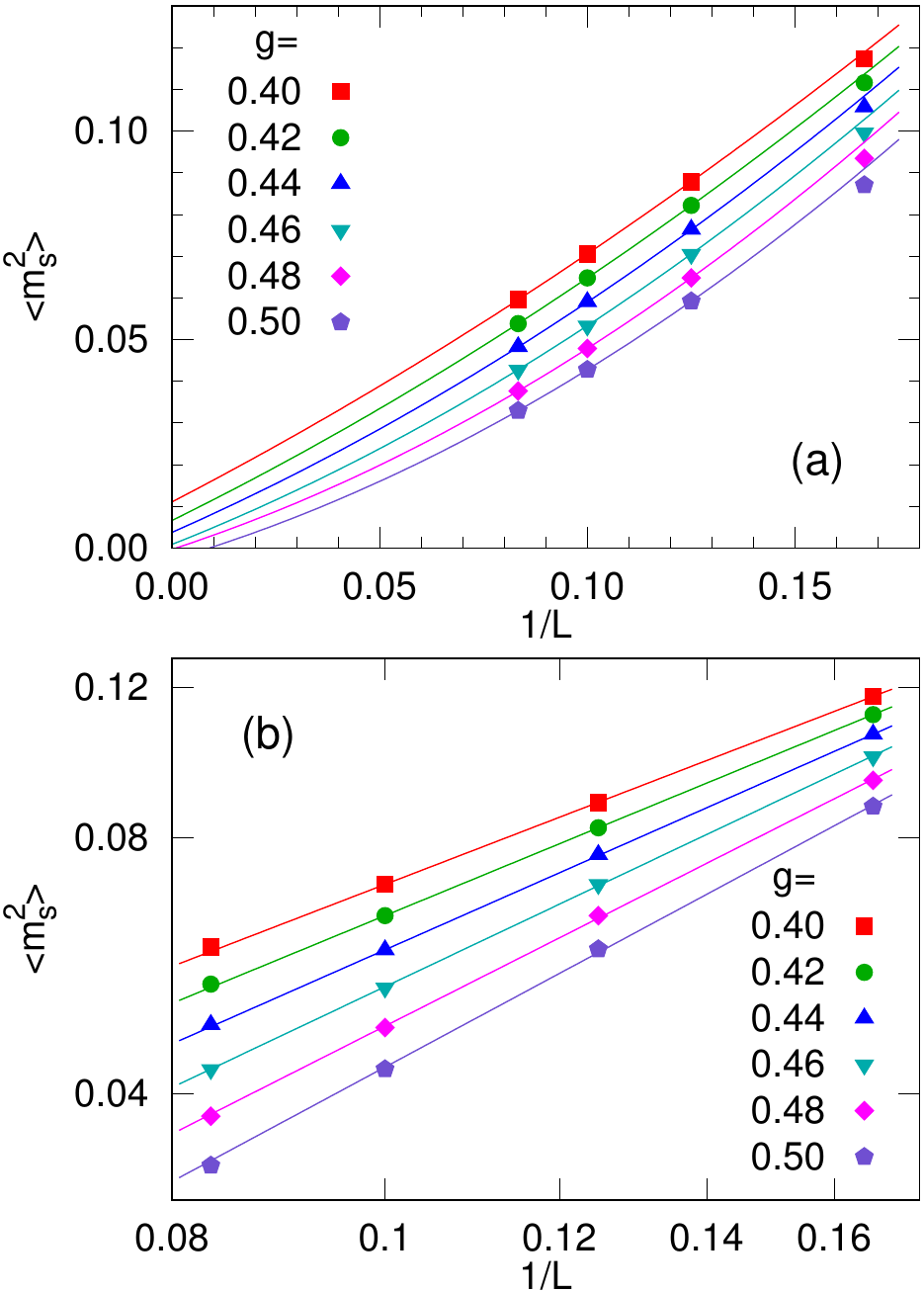}
\vskip-1mm
\caption{(a) Squared sublattice magnetization of the $J_1$-$J_2$
  model graphed vs $L^{-1}$. The curves are second order polynomials fitted to
  the $L=8,10,12$ data. (b) Log-log plot of the same data. Here the lines
  have been fitted to the $L=6,8,10$ points, to illustrate the upward curvature for larger sizes
  (here $L=12$) in the AFM ordered regime $g<0.46$.}
\label{j1j2stgm}
\end{center}
\vskip-2mm
\end{figure}

In Fig.~4 of the main paper we analyzed the sublattice magnetization of the $J_1$-$J_2$ model inside the 
putative SL phase and found power-law behaviors in the inverse system size. Here we present additional results
and analysis both below and above the crossing points $g_{c1} \approx 0.46$, demonstrating the existence of 
long-range AFM order for $g < g_{c1}$ and the absence of order for $g>g_{c1}$. Fig.~\ref{j1j2stgm}(a) shows
results graphed versus $L^{-1}$ together with second-order polynomial fitted to the data for $L=8,10,12$, representing
the expected asymptotic $L^{-1}$ form in the AFM state and a likewise expected $L^{-2}$ next correction. The curves extrapolate
to clearly positive values for $g=0.40,0.42$, and $0.44$, while the value at $g=0.46$ is almost zero. For larger
$g$ the extrapolated values are negative, indicating that the functional form used is incorrect. One should
expect the neglected higher-order corrections to also influence the extrapolated values for smaller $g$, and
the deviations between the fitted curve and data at $L=6$ give some indication of the size of the extrapolation 
errors for $g=0.40-0.46$. The results are consistent with the long-range order vanishing at $g\approx 0.46$, in 
excellent agreement with the result $g_{c1}\approx 0.46$ obtained from the singlet-quintuplet crossing points.
Cubic fits (not shown) to the $L\ge 6$ data result in slightly larger extrapolated values of $\langle m_s^2\rangle$,
but the $g$ dependence is less smooth than with the quadratic fits (likely reflecting sensitivity to the
small numerical errors in the individual data points and neglected corrections of still higher order). For $g\ge 0.46$ 
the cubic polynomials produce negative extrapolated values, supporting the conclusion drawn from the quadratic 
extrapolations that the AFM order vanishes close to $g=0.46$.

Further support for a critical AFM point at $g\approx 0.46$ is provided in Fig.~\ref{j1j2stgm}(b). Here we 
show the data on log-log scales, with straight lines (corresponding to power laws) drawn
through the $L=6,8,10$ data. For $g=0.40,0.42$, and $0.44$, the $L=12$ points fall above the lines, reflecting
an upward curvature as $L$ increases and AFM order is established. The behavior is similar to that of the 
$J$-$Q$ model in the AFM phase close to the critical point, e.g., at $J/Q=0.1$ in Fig.~\ref{jqstgm}.
For $g=0.46$, all four data points follow the fitted line very closely, while for larger $g$ the $L=12$ points fall 
below the fitted lines, reflecting negative curvature. In Fig.~4 of the main paper we fitted the data in the putative 
SL phase to a power law with an additional correction of higher power, required in order to fit all the available
data for $g > 0.46$.

Overall, these results and those in the main paper support a scenario of a critical
AFM--SL point at $g_{c1}\approx 0.46$ at which the scaling corrections are small, while for larger $g$
inside the SL phase the exponent of the asymptotic power law changes and corrections are needed to
explain the data on the relatively small systems accessible in DMRG calculations.

\subsection*{IV. Spin chains with long range interactions}

The spin-$1/2$ $J_1$-$J_2$ Heisenberg chain is a celebrated example of a system hosting
a quantum phase transition between quasi-long-range ordered (QLRO) and ordered
VBS phases. Defining $g=J_2/J_1$, the transition is located at $g_c\approx 0.2411$ \cite{nomura92}
and is accompanied by a critical level crossing of the lowest singlet and
triplet excitations. To study a quantum phase transition between a 1D long-range
AFM ordered and QLRO ground states, Laflorencie et al.~proposed \cite{laflorencie}
a Heisenberg chain with long-range interactions, with Hamiltonian
\begin{equation}
\label{1dlongrange}
H=\sum_{i=1}^L [\mathbf{S}_i\cdot\mathbf{S}_{i+1}+\lambda\sum_{r=2}^{L/2}J_r\mathbf{S}_i\cdot \mathbf{S}_{i+r}].
\end{equation}
where the couplings are of the form
\begin{equation}
J_r = \frac{(-1)^{r-1}}{r^{\alpha}},
\label{jrhberg}
\end{equation}
and $\alpha$ and $\lambda$ are both adjustable parameters.
Later on, to look for a possible 1D quantum phase transition between AFM
and VBS phases, a modification of the model was introduced in which the second
neighbor coupling $J_2$ changes sign, making it a frustrated term
\cite{sandvik10a};
\begin{equation}
\label{1dfrustratedlongrange}
H=\sum_{i=1}^L\sum_{r=1}^{L/2}J_r \mathbf{S}_i\cdot \mathbf{S}_{i+r},
\end{equation}
where the couplings are given by
\begin{equation}
J_2=g,\quad J_{r\neq 2}=\frac{(-1)^{r-1}}{r^{\alpha}}\left (1+\sum_{r=3}^{L/2}\frac{1}{r^{\alpha}}\right)^{-1},
\label{jrjq}
\end{equation}
where the adjustable parameters are $\alpha$ and $g$ and the normalization of $J_{r\not=2}$ is chosen
such that the sum of all nonfrustrated ($r\not=2$) interactions $|J_r|$ equals $1$.

For the unfrustrated chain, a curve of continuous AFM--QLRO transitions was mapped out in
the $(\alpha,\lambda)$ plane \cite{laflorencie}. In the frustrated chain, it was found that, by fixing
the frustration strength $g$ and tuning the exponent $\alpha$ controlling the long-range interaction,
two quantum phase transitions take place along this path \cite{sandvik10a}; a QLRO-VBS transition
with singlet-triplet excitation level crossing as in the $J_1$-$J_2$ chain, and, for smaller $\alpha$,
an AFM-QLRO transition accompanied by another level crossing. This second crossing was claimed to
be a singlet-singlet crossing, but it turns out that (as found in the course of the work reported here)
that the total spin $S$ of one of the levels was misidentified as a singlet though it actually is
an $S=2$ quintuplet and the crossing discussed is a singlet-quintuplet crossing. In other
respects we fully agree with the previous results. Thus, the
behavior of the frustrated long-range interacting chain upon increasing $\alpha$
is very similar to that we have observed in the square-lattice $J_1$-$J_2$ model
upon increasing $g=J_2/J_1$.

In the case of the unfrustrated model with $J_r$ given by Eqs.~(\ref{jrhberg}) we also 
expect the AFM-QLRO quantum phase transition to be accompanied by a singlet-quintuplet
excitation crossing, though level crossings were not discussed in Ref.~\onlinecite{laflorencie}.
Here we revisit the quantum phase transitions in both the frustrated and unfrustrated
chain models, analyzing level crossings obtained by the SU(2) DMRG method to push to
large system sizes than what was possible with the previous Lanczos calculations
in Ref.~\onlinecite{sandvik10a}. This will provide us with further,
indisputable evidence that the AFM--QLRO transition in the 1D system indeed is
accompanied with a singlet-quintuplet crossing. This in turn gives added credence
to our claim of this scenario for the 2D $J_1$-$J_2$ model.

In the unfrustrated model we set $\lambda=1$ in Eq.~(\ref{1dlongrange}) and study the AFM-QLRO
quantum phase transition by tuning the long-range interaction exponent $\alpha^{-1}$. In
the frustrated  model, Eq.~(\ref{1dfrustratedlongrange}), we choose a path with fixed $g=0.3$
in Eq.~(\ref{jrjq}) and vary $\alpha^{-1}$ from $1$ to $0$, thus passing through both the AFM--QLRO
and QLRO--VBS transitions.

\begin{figure}[t]
\begin{center}
\includegraphics[width=6.5cm]{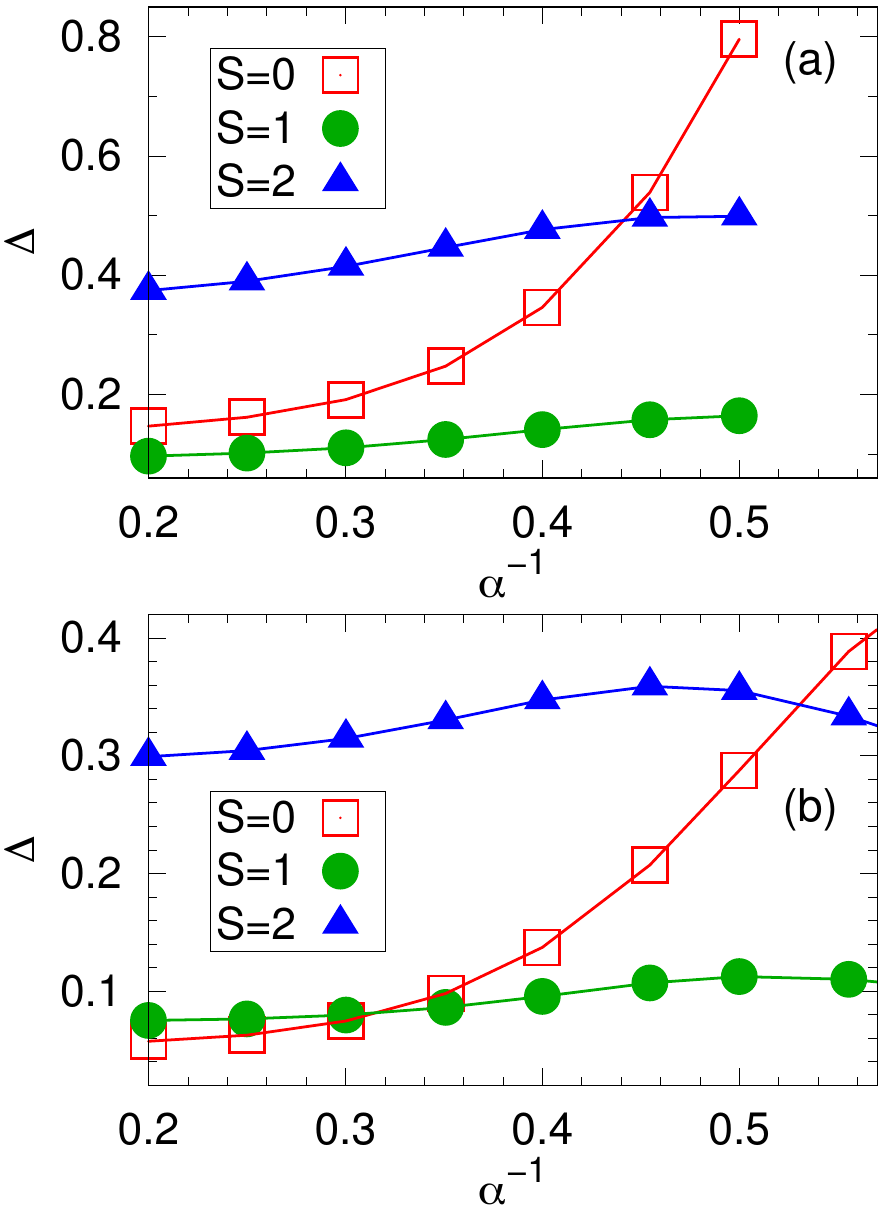}
\vskip-1mm
\caption{The lowest excitation gaps in singlet ($S=0$), triplet ($S=1$) and
  quintuplet ($S=2$) sectors in (a) the unfrustrated spin chain with long range
  interactions and in (b) the frustrated spin chain with long range
  interactions, both for chain length $L=48$. The gaps are graphed vs the exponent $\alpha^{-1}$ controlling
  the decay of the long-range interaction with (a) $\lambda=1$ in Eq.~(\ref{1dlongrange}) and
  (b) $g=0.3$ in Eq.~(\ref{jrjq}). Solid lines are guides to the eye.}
\label{levelcrosschain}
\end{center}
\vskip-2mm
\end{figure}

\begin{figure}[t]
\begin{center}
\includegraphics[width=6.5cm]{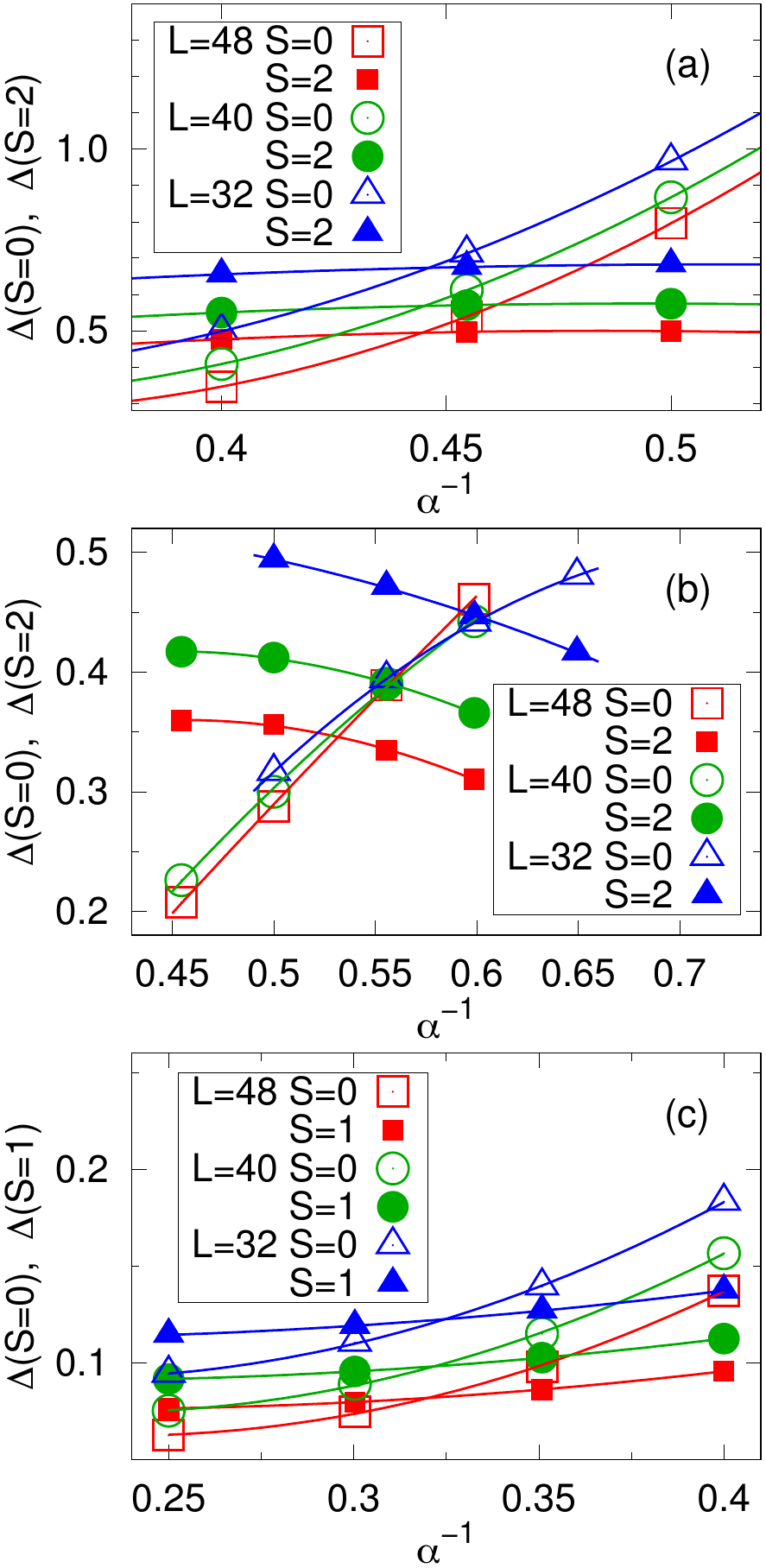}
\vskip-1mm
\caption{Gaps vs $\alpha^{-1}$ in (a) the unfrustrated spin chain with
  long-range interactions at $\lambda=1$ in Eq.~(\ref{1dlongrange}) and (b-c)
  the frustrated spin chain with long range interactions at $g=0.3$ in Eq.~(\ref{jrjq}).
  All three crossing points $g_{c}(L)$ (singlet-quintuplet, unfrustrated model),
  $g_{c1}(L)$ (singlet-quintuplet, frustrated model) and $g_{c2}(L)$
  (singlet-triplet, frustrated model) are extracted using second-order
  polynomial fits (the curves shown).}
\label{levelcrosschain2}
\end{center}
\vskip-2mm
\end{figure}

In Fig.~\ref{levelcrosschain} we plot the lowest singlet, triplet and
quintuplet gaps of $L=48$ chains versus $\alpha^{-1}$ at (a) fixed $\lambda=1$
in the unfrustrated model and (b) fixed $g=0.3$ in the frustrated chain.
In both models, the crossing points of the lowest
singlet-quintuplet excitations indicate the AFM-QLRO quantum phase transitions,
based on the behaviors previously found for the sublattice magnetization.
In the frustrated case, the crossing of the lowest singlet and triplet excitations
marks the QLRO-VBS quantum phase transition, in analogy with the case of the
conventional $J_1$-$J_2$ Heisenberg chain without the $J_{r>2}$ (which also
corresponds to $\alpha=\infty$ in the long-range model).

\begin{figure}[t]
\begin{center}
\includegraphics[width=6.5cm]{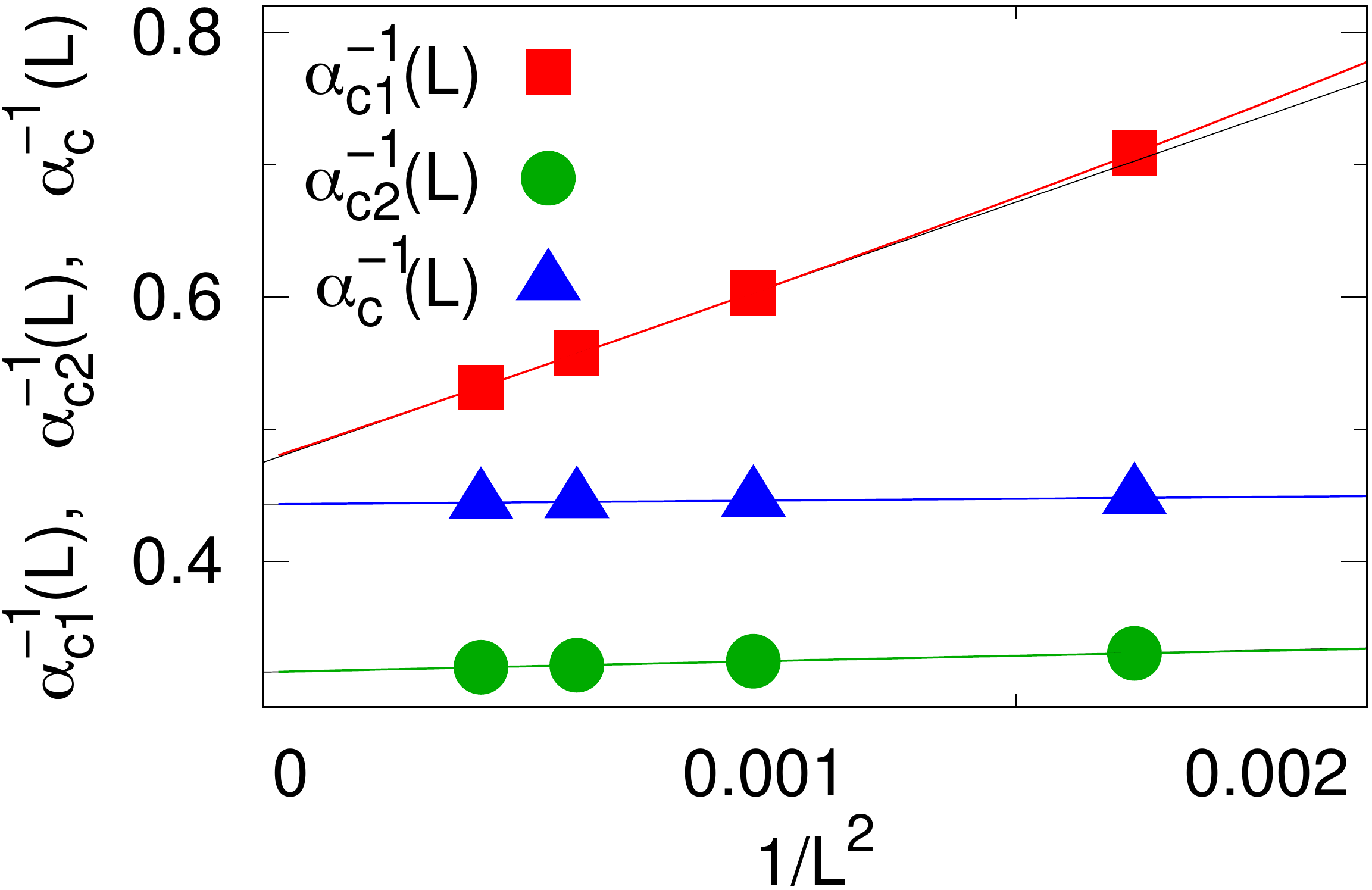}
\vskip-1mm
\caption{The gap-crossing points in Fig.~\ref{levelcrosschain2} graphed vs
  $L^{-2}$. The singlet-quintuplet crossing points in the unfrustrated case
  are shown as blue triangles, the singlet-quintuplet points in the
  frustrated case as red squares, and the singlet-triplet point in the
  frustrated case as green circles. The black lines are drawn through the
  $L = 40$ and $48$ points, while the corresponding colored curves are of the
  form
  $\alpha^{-1}_{c}(L) = \alpha^{-1}_{c}(\infty) + aL^{-2}(1 + bL^{-\omega})$
  with $\alpha^{-1}_{c}(\infty) \approx 0.4434$,
  $\alpha^{-1}_{c1}(\infty) \approx 0.476$, and
  $\alpha^{-1}_{c2}(\infty) \approx 0.316$. }
\label{gc_fit_chain}
\end{center}
\vskip-2mm
\end{figure}

We further examine the drifts of these critical level crossings for different system
sizes, $L=32,40,48$, in the critical regions. In Fig.~\ref{levelcrosschain2} the gaps
are fitted to second order polynomials to interpolate the finite-size critical points
$\alpha^{-1}_{c}(L)$ (singlet-quintuplet in the unfrustrated case), $\alpha^{-1}_{c1}(L)$
(singlet-quintuplet in the frustrated case), and $\alpha^{-1}_{c2}(L)$ (singlet-triplet
in the frustrated case). Fig.~\ref{gc_fit_chain} shows the size dependence of all these
crossing points versus $L^{-2}$ along with lines drawn through the data for the largest
two sizes, $L=40$ and $48$. We also show fitted curves including a higher-order
correction, which give the infinite-size extrapolated values $\alpha^{-1}_{c}=0.4434$,
$\alpha^{-1}_{c1}=0.476$, and $\alpha^{-1}_{c2}=0.316$. In the unfrustrated model, the critical
value $\alpha^{-1}_{c}=0.4434$, i.e., $\alpha_c=2.255$, is fully consistent with the quantum
critical point $\alpha_c=2.225\pm 0.025$ found by analyzing QMC results for the AFM order
parameter in Ref.~\cite{laflorencie}. Thus, there is no doubt that the singlet-quintuplet
crossing really marks the AFM--QLRO transition in the unfrustrated chain and there
is no reason why this should not be the case also in the frustrated model; indeed the
behavior of the order parameters (not shown here) also supports the existence
of the phase transition.

\subsection*{V. Singlet-singlet level crossing}

As seen in Fig.~2 in the main paper, there is also a singlet-singlet level crossing in the neighborhood of 
the singlet-quintuplet point analyzed in the main paper. We call the singlet-singlet crossing point 
$g_{c1}^{\prime}(L)$ and investigate its behavior here.

\begin{figure}[t]
\begin{center}
\includegraphics[width=6.5cm]{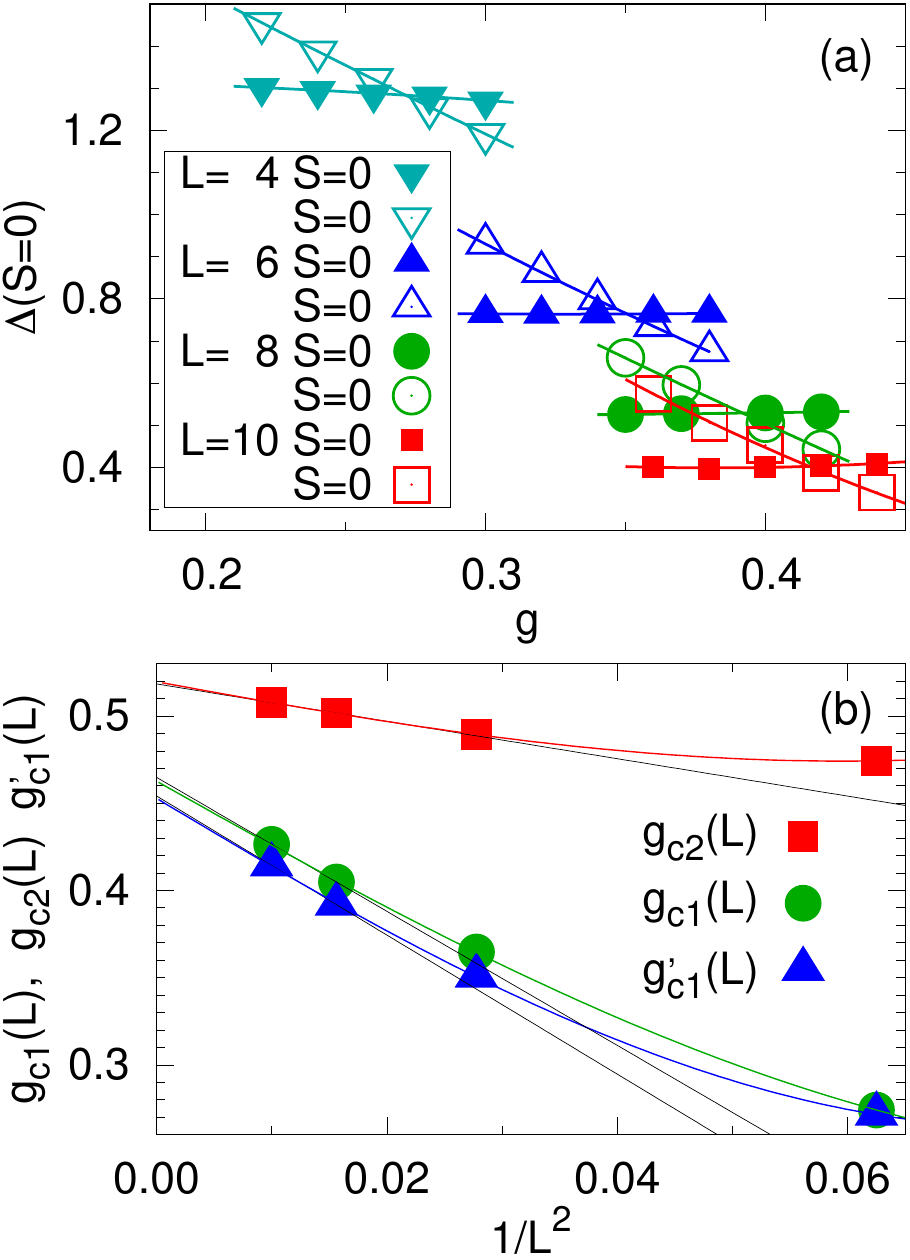}
\vskip-1mm
\caption{(a) The singlet gaps graphed in Fig.~2 in the main paper for system sizes $L=4,6,8,10$
in the neighborhood of the points at which the two levels cross each other. The crossing
points are extracted using second-order polynomial fits. (b) The singlet-singlet 
crossing points in (a) graphed vs $L^{-2}$ (blue triangles), together with the singlet-quintuplet 
crossing points (green circles) and singlet-triplet crossing points (red squares). The  black line 
following the blue triangles is drawn through the $L=8,10$ points, while the blue curve is of the form
$g_c(L) = g_c(\infty) + aL^{-2}(1+bL^{-\omega})$ with $g^{\prime}_{c1}(\infty) \approx 0.453$, and
$\omega^{\prime}_1 \approx 3.7$ for fitting the singlet-singlet crossing points.}
\label{sscross}
\end{center}
\vskip-2mm
\end{figure}

In Fig.~\ref{sscross}(a) we demonstrate the singlet-singlet level crossing for
different system sizes and study the trend of this crossing point as a function
of the inverse system size in Fig.~\ref{sscross}(b). A plausible $L^{-2}$ correction
is again assumed here. Then a rough extrapolation to infinite size by a line drawn
through the $L=8$ and $L=10$ points in the figure gives $g^{\prime}_{c1}\approx 0.454$.
On including a correction with the same fitting form as in the singlet-triplet case,
the extrapolated value moves slightly down to $g^{\prime}_{c1}=0.453$. This value is
very close to $g_{c1}=0.463$, marking the AFM-SL ground states phase transition as 
given by the singlet-quintuplet crossing point. Thus, it seems plausible that the 
AFM-SL transition is associated with both singlet-singlet and singlet-quintuplet 
excitation crossings, though larger system sizes would be needed to confirm whether
the points really flow to the same values. 

It should be noted that we have not found 
any singlet-singlet crossing at the AFM--QLRO transition in the case of the 1D chain 
discussed above in Sec.~III. The singlet-quadruplet crossing point along with its
scaling in energy as $L^{-1}$, shown in Fig.~3(b) of the main paper, is also a more
clear-cut indicator of a transition out of the AFM state in the sense that we know
that the $S=2$ level is a quantum-rotor state that scales as $L^{-2}$ in the AFM 
state. In principle, the singlet-singlet crossing could be accidental and unrelated 
to the AFM--SL transition, though the close proximity to the singlet-quadruplet crossing 
in our extrapolations based on rather small sizes would suggest that it actually is 
also associated with the transition in the 2D model.
\end{document}